\documentclass[twocolumn, amssymb, superscriptaddress, aps]{revtex4-2}
\usepackage{graphicx}
\usepackage[T1]{fontenc}
\usepackage{lmodern}
\usepackage{color}
\usepackage{natbib}
\usepackage{amsmath}
\usepackage{multirow} 
\usepackage{lipsum}

\begin{document}

\title{Unconventional quantum oscillations and evidence of nonparabolic electronic states in quasi-two-dimensional electron system at complex oxide interfaces}
\author{Km Rubi}
\email{Corresponding author: rubi@lanl.gov}
\affiliation {National High Magnetic Field Laboratory, Los Alamos National Laboratory, Los Alamos, New Mexico 87544 USA}
\affiliation{High Field Magnet Laboratory (HFML-EMFL) and Institute for Molecules and Materials, Radboud University, 6525 ED Nijmegen, The Netherlands}

\author{{Denis R. Candido}}
\affiliation {{Department of Physics and Astronomy, University of Iowa, Iowa City, Iowa 52242 USA}}

\author{Manish Dumen}
\affiliation{Quantum Materials and Devices Unit, Institute of Nano Science and Technology, Mohali, Punjab 140306, India}

\author{Shengwei Zeng}
\email{Present address: Institute of Materials Research and Engineering (IMRE), Agency for Science, Technology and Research (A*STAR), 2 Fusionopolis Way, Innovis \#08-03, Singapore 138634, Republic of Singapore}
\affiliation{Department of Physics, National University of Singapore, 117551 Singapore}

\author{Emily L.Q.N. Ammerlaan}
\affiliation{High Field Magnet Laboratory (HFML-EMFL) and Institute for Molecules and Materials, Radboud University, 6525 ED Nijmegen, The Netherlands}

\author{Femke Bangma}
\affiliation{High Field Magnet Laboratory (HFML-EMFL) and Institute for Molecules and Materials, Radboud University, 6525 ED Nijmegen, The Netherlands}

\author{Mun K. Chan}
\affiliation {National High Magnetic Field Laboratory, Los Alamos National Laboratory, Los Alamos, New Mexico 87544 USA}

\author{Michel Goiran}
\affiliation{Laboratoire National des Champs Magnétiques Intenses (LNCMI-EMFL), Université de Toulouse, CNRS, INSA, UPS, 143 Avenue de Rangueil, 31400 Toulouse, France}

\author{Ariando Ariando}
\affiliation{Department of Physics, National University of Singapore, 117551 Singapore}

\author{Suvankar Chakraverty}
\affiliation{Quantum Materials and Devices Unit, Institute of Nano Science and Technology, Mohali, Punjab 140306, India}

\author{Walter Escoffier}
\affiliation{Laboratoire National des Champs Magnétiques Intenses (LNCMI-EMFL), Université de Toulouse, CNRS, INSA, UPS, 143 Avenue de Rangueil, 31400 Toulouse, France}

\author{Uli Zeitler}
\affiliation{High Field Magnet Laboratory (HFML-EMFL) and Institute for Molecules and Materials, Radboud University, 6525 ED Nijmegen, The Netherlands}

\author{Neil Harrison}
\affiliation {National High Magnetic Field Laboratory, Los Alamos National Laboratory, Los Alamos, New Mexico 87544 USA}

\date{\today}

\begin{abstract}
The simultaneous occurrence of electric-field controlled superconductivity and spin-orbit interaction makes two-dimensional electron systems (2DES) constructed from perovskite transition metal oxides promising candidates for the next generation of spintronics and quantum computing. It is, however, essential to understand the electronic bands thoroughly and verify the predicted electronic states experimentally in these 2DES to advance technological applications. Here, we present novel insights into the electronic states of the 2DES at oxide interfaces through comprehensive investigations of Shubnikov-de Haas oscillations in three different systems: EuO/KTaO$_3$, LaAlO$_3$/SrTiO$_3$, and  amorphous-LaAlO$_3$/KTaO$_3$. To accurately resolve these oscillations, we conducted transport measurements in high magnetic fields up to 60 T and low temperatures down to 100 mK. For 2D confined electrons at these interfaces, we observed a progressive increase of oscillations frequency and cyclotron mass with the magnetic field. We interpret these universal and intriguing findings by considering the existence of non-trivial electronic bands, for which the $E-k$ dispersion incorporates both linear and parabolic relations. In addition to providing experimental evidence for {non-parabolic} electronic states in KTaO$_3$ and SrTiO$_3$ 2DES, the unconventional oscillations presented in this study establish a new paradigm for quantum oscillations in 2DES based on perovskite transition metal oxides, where the oscillation frequencies in $1/B$ exhibit quadratic dependence on the magnetic field. 
 \end{abstract}

\maketitle

\section{Introduction}

Two-dimensional electron systems (2DES) have been observed at the surface and interface of many perovskite transition metal oxides, so-called complex oxides. Particularly, widely studied 2DES based on SrTiO$_3$ (STO) and KTaO$_3$ (KTO) exhibit various intriguing phenomena, including a large magnetoresistance, Rashba spin-orbit interaction \cite{PhysRevLett.104.126803, PhysRevLett.108.117602, wadehra2020planar}, 2D superconductivity \cite{li2011coexistence, bert2011direct, ueno2011discovery, chen2021two, liu2021two}, and magnetism \cite{zhang2018high}, which do not exist in their bulk counterparts. 
The coexistence of these phenomena gives these systems a multi-functional character, with potential applications in spintronics \cite{noel2020non, vicente2021spin} as well as in the field of topological quantum computing \cite{chung2016dislocation, barthelemy2021quasi}. However, a comprehensive understanding of the electronic structure that gives rise to these interesting phenomena remains elusive. {Furthermore, these systems offer a unique opportunity to explore the physics of 2D-confined $d$ electrons, distinct from the scenario of the 2DESs in conventional III-V semiconductor heterostructures.} 

STO and KTO based 2DES exhibit several similarities in terms of their calculated band structures. For example,  the electrons occupy crystal-field split $t_{2g}$ orbital of $d$ bands ($3d$ for STO and $5d$ for KTO), and the combination of 2D confinement and spin-orbit interactions gives rise to multiple bands with mixed orbital characters of $d_{xy}$, $d_{xz}$, and $d_{yz}$ due to the avoided crossing between light ($d_{xy}$) and heavy ($d_{xz}$/$d_{yz}$) subbands. Heeringen \emph{et al.} \cite{PhysRevB2013} predicted strongly anisotropic nonparabolic subbands for 2DES at the LaAlO$_3$(LAO)/STO interface. Furthermore, {non-trivial} topological states with linear dispersion are predicted for STO and KTO 2DES in the vicinity of avoided crossing points in $\Gamma$-M direction of the first Brillouin zone \cite{vivek2017, vaz2019, kakkar2023}. While experiments based on the Shubnikov-de Haas (SdH) effect and angle-resolved photoemission spectroscopy (ARPES) have verified the existence of several subbands of different effective masses for both STO \cite{mccollam2014quantum, meevasana2011creation, rodel2016universal} and KTO 2DES \cite{rubi2021electronic, PhysRevLett.108.117602, santander2012orbital}, the signature of nonparabolic subbands or topological states in these systems have not yet been perceived through experiments. Interestingly, the STO-2DES exhibits peculiar SdH oscillations that are not periodic in inverse magnetic field \cite{fete2014large, MingYang2016, trier2016quantization, cheng2018shubnikov, rubi2020aperiodic}. The aperiodicity in oscillations perceived in high magnetic fields has been tentatively attributed to different mechanisms (e.g., Rashba spin-orbit interaction \cite{fete2014large, MingYang2016}, Zeeman splitting \cite{trier2016quantization}, magnetic depopulation of magnetoelectric subbands \cite{cheng2018shubnikov}, and magnetic field-induced change in carrier density \cite{rubi2020aperiodic}) in different investigations; and its physical origin has not yet reached to a consensus. Furthermore, despite a comparable electronic band structure to the STO-2DES, the existence of aperiodic SdH oscillations in KTO-2DES remains unclear from previous studies \cite{harashima2013coexistence, kumar2021observation, rubi2021electronic, yan2022ionic}.  

In our quest to unravel the origin of aperiodic quantum oscillations and uncover {peculiar electronic} states in {$d$-electrons 2D systems}, we conducted a thorough experimental investigation of the SdH oscillations at the interfaces of EuO/KTO, LAO/STO, and {amorphous(\emph{a})-LAO/KTO.} In order to capture the oscillations with utmost precision, we measured electrical transport in high magnetic fields, utilizing both a pulsed field (60 T) and a dc field (35 T), and at ultra-low temperatures (as low as 0.1 K). {For clarity, we first compare the experimental results of EuO/KTO and LAO/STO in section III, with the discussion on a-LAO/KTO presented later. However, we consolidate the key findings from all three interfaces in section IV.} 
By examining the tilt-angle dependence of the quantum oscillations for EuO/KTO and LAO/STO, we reveal the presence of itinerant electrons that are confined in the 2D interface region, coexisting with the carriers that disperse deeper into the STO and KTO. Interestingly, we observed that {all three} interfaces exhibit a progressive increase in the effective cyclotron mass, estimated from the SdH oscillations, as well as an apparent increase of the oscillations frequency in {$1/B$} with increasing magnetic field. Notably, we found that the increase in cyclotron mass follows an almost linear trend, while the change in frequencies exhibits a quadratic relationship with the magnetic field. We attribute this behavior to the presence of non-trivial electronic bands {determined by spin-orbit interaction,} wherein energy dispersion in $k$-space incorporates both linear and quadratic terms. These findings offer valuable insights into the distinct electronic properties and subband structure at the interfaces of complex oxides.

\section{Methods}
As depicted in Fig.1(a) and (b), the EuO/KTO sample consists of a 10 nm thin film of EuO on KTO (001) substrate, while LAO/STO is made of $\sim$ 3.2 nm (8 u.c.) thin film of LAO on STO (001) substrate. Both KTO and STO substrates are 0.5 mm thick. We used a pulsed laser deposition technique to grow EuO and LAO thin films. For the LAO/STO sample, a mask of amorphous AlN was deposited on STO before LAO growth to obtain a Hall-bar patterned sample. One can find the growth details for EuO/KTO in Ref. \cite{kumar2021observation}  and for LAO/STO in Ref. \cite{rubi2020aperiodic}. 
We carried out longitudinal and Hall resistance measurements simultaneously on the EuO/KTO sample in high pulsed magnetic fields ($B_{max}$  = 60 T and pulse time $\sim$ 25 ms) and down to the temperature of 0.5 K in the $^3$He system. To achieve a high signal-to-noise ratio {from EuO/KTO measurements} in a pulsed field, we used an excitation current of amplitude 30 $\mu$A and frequency up to 256 kHz. {To rule out the possibility of Joule heating from the 30 $\mu$A  current, we compare the data measured by applying lower and higher current at a fixed frequency and temperature (Appendix A1)}. We measured LAO/STO in a high continuous magnetic field ($B_{max}$ = 35 T) and at low temperatures down to 0.1 K in a dilution fridge by applying a quasi dc excitation of 0.1 $\mu$A. {The a-LAO/KTO Hall-bar patterend sample was measured in high pulsed fields up to 55 T (pulse time $\sim$ 300 ms) using a dc excitation current of 1 $\mu$A}. The measurements at different tilt angles were performed using in-situ sample rotators devised explicitly for the dilution fridge and the $^3$He fridge used in the extreme environment of high magnetic fields. To probe the interface using transport measurements, we made electrical contacts for all samples using a wire bonder. In particular,  we measured an unpatterned EuO/KTO sample at the lowest temperature in both up and down field directions (for details see Fig. A3(a) in Appendix A1), and attained the {antisymmetric $R_{yx}$ using the formulas $R_{yx} = \frac{R_{yx}(B\uparrow)-R_{yx}(B\downarrow))}{2}$ for a carrier density estimation. To have a precise estimation of mobility, we measured zero-field sheet resistance of EuO/KTO using the Van der Pauw method (Appendix A2).}

\section{Experimental results}
\subsection{Electrical properties and quantum oscillations}
\begin{figure}[!htp]
\includegraphics[width=8.7cm]{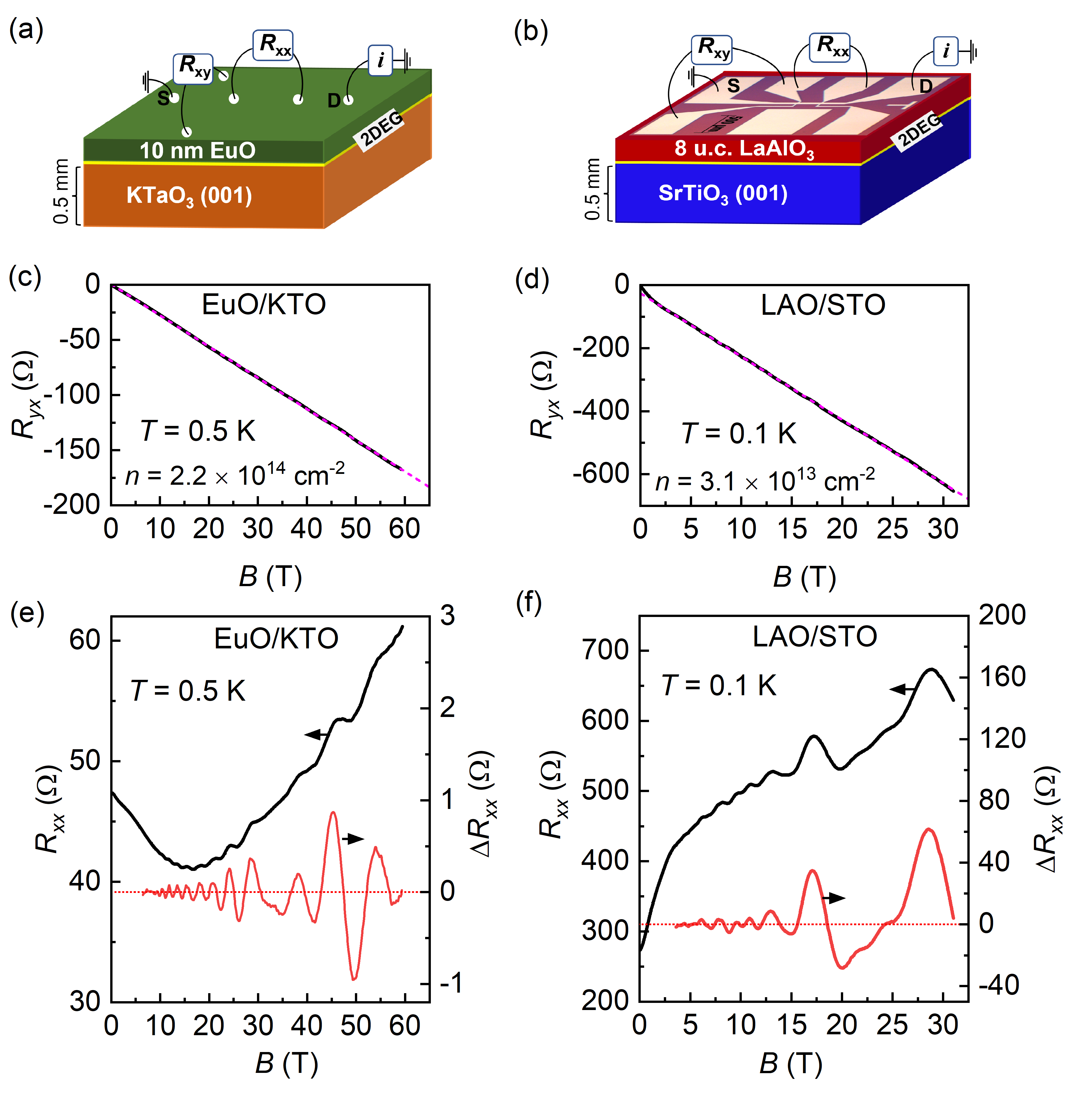} 
\caption{\label{F1} Schematics of EuO/KTO and LAO/STO heterostructures and their magnetotransport data. Left column: EuO/KTO and right column: LAO/STO. (a) and (b) schematic diagrams with electrical contacts and transport measurement scheme. (c) and (d) Hall-resistance $R_{yx}$($B$) data (solid line) and linear fit (dashed line) to it in high-field regime (> 5 T) extended to zero-field. (e) and (f) Left y-axis: Magnetic field dependence of longitudinal resistance, $R_{xx}$($B$). Right y-axis: Oscillating resistance, $\Delta R_{xx}$, estimated by subtracting the fit data from the measured data. The dotted horizontal lines display the zero value of $\Delta R{xx}$. {For EuO/KTO, the  $R_{yx}$($B$) is antisymmetrized from the measurements in up and down field directions for estimating carrier density, however,  $R_{xx}$($B$) is the data measured in up field direction.}}
\end{figure}

\begin{figure*}[!htp]
\includegraphics[width=6.5in]{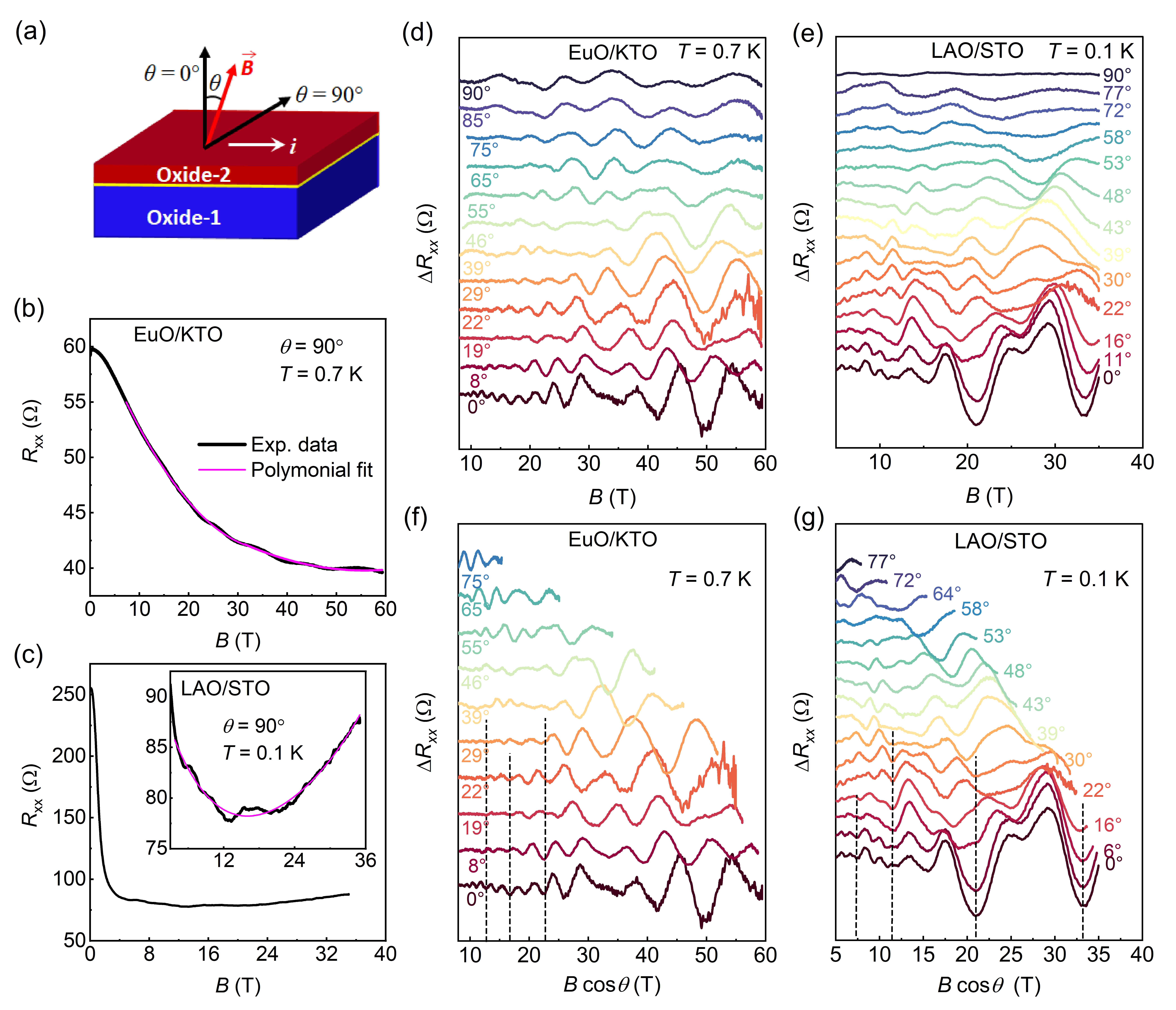}
\caption{\label{F2} Angular dependence of quantum oscillations for EuO/KTO and LAO/STO.
(a) Illustration of tilt angle $\theta$, i. e. the angle between the normal to the interface and the magnetic field direction. Current is perpendicular to the magnetic field for all $\theta$ values. (b) and (c) $R_{xx}$ ($B$) at magnetic fields oriented parallel to the interface, $\theta$ = $90^{\circ}$, for EuO/KTO and LAO/STO, respectively. Unlike at $\theta$ = $0^{\circ}$, both systems show negative magnetoresistance at  $\theta$ = $90^{\circ}$. Inset of (c): a zoom-in view of the $R_{xx}$ ($B$) in high magnetic fields for LAO/STO. (d) and (f) Quantum oscillations at different $\theta$ values for EuO/KTO plotted as a function of $B$ and $B_{\perp} = B$cos($\theta$), respectively. (f) and (g) Quantum oscillations for various $\theta$ values for LAO/STO. The dash lines are guide to the eyes.}
\end{figure*}

\begin{figure*}[!htp]
\includegraphics[width=6.5in]{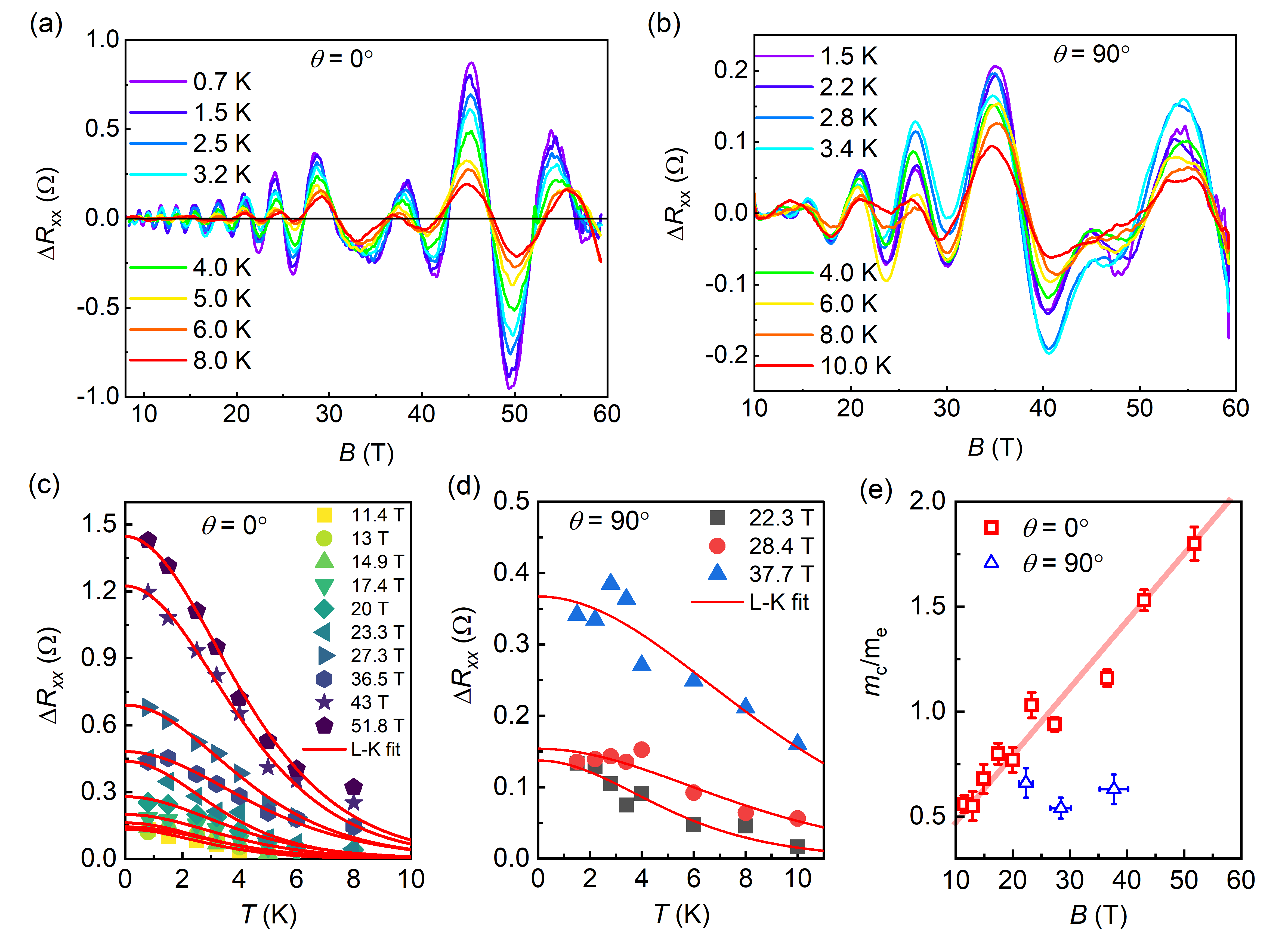}
\caption{\label{F2} Cyclotron mass analysis for EuO/KTO.
 $\Delta R_{xx}$ at different temperatures for (a) $\theta = 0^\circ$ and (b) $\theta = 90^\circ$. Temperature dependence of oscillations amplitude (symbols) with fit (solid line) to Lifshitz-Kosevick theory \emph{i.e.} Eq. (1) for (c) $\theta = 0^\circ$ and (d) $\theta = 90^\circ$. (e) A comparison of cyclotron mass at $\theta = 0^\circ$ and $90^\circ$estimated from the fit to Eq. (1). The solid line is guide to the eyes for $\theta = 0^\circ$ data.}
\end{figure*}

Fig.1 (c) and (d) show the Hall resistance $R_{yx}$($B$) for EuO/KTO and LAO/STO interfaces, respectively, measured at the lowest temperature possible for each case and in magnetic fields ($B$) oriented perpendicular to the interface. Except in the low field regime of 0 - 4 T, the $R_{yx}$($B$) is linear for both interfaces. From the slope of the linear fit to the $R_{yx}$($B$) for $B$ > 5 T, we estimate the carrier density of 2.2 $\times 10^{14}$ cm$^{-2}$ for EuO/KTO and 3.1 $\times$ $10^{13}$ cm$^{-2}$ for LAO/STO. 
However, despite having a lower effective mass \cite{kumar2021observation}, the carriers at the EuO/KTO interface exhibit a lower Hall mobility $\sim$ {650} cm$^2$V$^{-1}$s$^{-1}$ than thereof  LAO/STO $\sim$ 2350 cm$^2$V$^{-1}$s$^{-1}$. We attribute the lower carrier mobility in EuO/KTO to the {spin scattering of charge carriers \cite{zhang2018high} and the high density of carriers. Different from LAO/STO, the electrons in EuO/KTO are introduced via Eu substitution on K site or oxygen vacancies \cite{liu2021two, yan2022ionic}, which also act as a potential scattering center and restrict the mobility of charge carriers.}

The left y-axes of Fig.1(e) and (f) display the magnetic field dependence of longitudinal resistance  $R_{xx}$($B$) for LAO/STO and EuO/KTO, respectively. {Both interfaces exhibit positive magnetoresistance (MR) under the field applied perpendicular to the interface. The distinct dependence of MR on the magnetic field – such as being quadratic in the low-field regime and quasi-linear in the high-field regime – suggests the presence of different mechanisms, such as Lorentz force, interface scattering \cite{wang2011, li2022} or inhomogeneity\cite{niu2021}, responsible for MR in these distinct field regimes (Appendix A3).} For both interfaces, the quantum oscillations originating from the quantization of closed cyclotron orbits are superimposed on a positive MR. We show the oscillating resistance $\Delta R_{xx}$ after subtracting a smooth background on the right y-axes of each panel. For both interfaces, the non-monotonic enhancement of the oscillations amplitude with increasing magnetic field indicates the presence of more than one frequency, as verified by multiple peaks in the Fast Fourier Transform (FFT), which will be discussed in detail later. A two-order of magnitude smaller amplitude of the quantum oscillations in EuO/KTO than LAO/STO is consistent with a lower mobility of carriers at the EuO/KTO interface. 

\subsection{Tilt-angle dependence of quantum oscillations}

To examine the dimensionality of the electron systems at the EuO/KTO and LAO/STO interfaces, we measured both samples at different tilt angles ranging from $0^\circ$ to $90^\circ$. The tilt angle $\theta$, as illustrated in Fig. 2(a), is defined as the angle between the magnetic field $B$ and the normal to the interface. For all field orientations, $B$ is perpendicular to the current. First, it is worth mentioning that both interfaces show a large negative MR for the in-plane field orientation, $\theta = 90^\circ$, as shown in the main panels of Fig.2(b) and 2(c). The magnitude of the negative MR is larger for LAO/STO (see Appendix A3), even though this system does not consist of any magnetic material, which could induce magnetic proximity effect on the interfacial conducting sheets, as reported for EuO/KTO \cite{zhang2018high}. The negative MR in complex oxide interfaces with the application of an in-plane magnetic field can be attributed to the combined effect of spin-orbit coupling and long-range impurity scattering \cite{PhysRevLett.115.016803}. Additionally, the positive MR for LAO/STO in the high magnetic fields (inset of Fig. 2(c)) can be explained by the domination of conventional orbital MR in this regime as the higher carriers mobility in LAO/STO leads to the completion of more cyclotron orbits compared to EuO/KTO.

After subtracting a smooth background from $R_{xx}$($B$) measured at different tilt angles, we show $\Delta R_{xx}$ as a function of the $total$ magnetic field in Fig.2(d) and (e) for EuO/KTO and LAO/STO, respectively. Both systems show a complex shift in oscillations' minima and maxima positions at least up to $\theta$  = $45^{\circ}$. We, however, do not perceive a noticeable change in oscillations for $\theta > 65^{\circ}$. For both interfaces, the fixed quantum oscillations pattern in the regime of $\theta = 75^{\circ} - 90^{\circ}$, as verified by FFT analysis in Appendix A4, provides evidence for the coexistence of three-dimensional conduction channels. 

To identify the two-dimensional (2D) nature of the electron systems, we plot $\Delta R_{xx}$ of EuO/KTO and LAO/STO as a function of the perpendicular component of magnetic field, $B$cos($\theta$), in Fig. 2(f) and (g), respectively. For EuO/KTO, the low-field oscillations follow a cos($\theta$) scaling for $\theta < 30^{\circ}$, indicating the 2D confinement of conduction electrons at the interface. However, on comparing Fig. 2(d) and (e), we find the high field oscillations (> 25 T) to follow a scaling of neither $B_{total}$ nor $B$cos($\theta$), indicating the superposition of oscillations originating from 2D and 3D Fermi surfaces. In contrast, the oscillations in LAO/STO exhibit a $B$cos($\theta$) scaling (depicted by vertical dashed lines) up to the high fields (35 T), except a few minima that might be affected by the crossover of Landau levels of multiple electronic subbands. Overall, both samples reveal a 2D confinement of electrons at the interface, along with a fraction of electrons dispersed deep into KTO and STO. Moreover, the existence of quantum oscillations in Hall resistance $R_{xy}(B)$ and the angular dependence of $R_{xy}(B)$ and $R_{xx}(B)$ suggest that the majority of electrons are confined in 2D (Appendix A5).

\subsection{Magnetic field dependence of cyclotron mass}

\begin{figure}[!htp]
\includegraphics[width=3.4in]{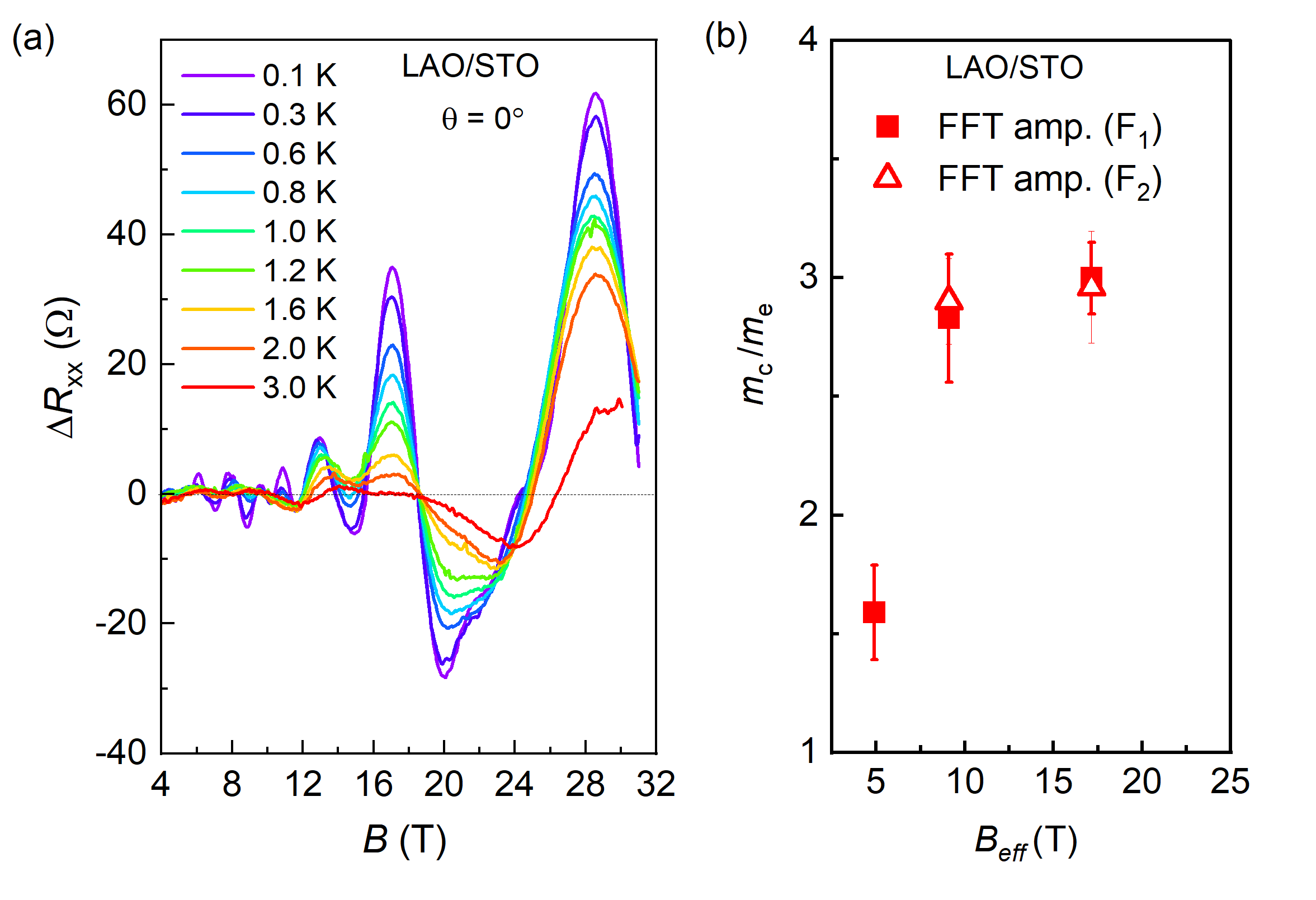}
\caption{\label{F2} Cyclotron mass in the 2DES at LAO/STO interface. (a) Magnetic field dependence of $\Delta R_{xx}$ at different temperatures ranging from 0.1 K to 3.0 K. (b) Cyclotron mass estimated from the L-K fit to the FFT amplitude. The FFT spectra at different temperatures and the L-K fit to the FFT amplitude are displayed in Appendix A6, Fig. A9.}
\end{figure}

\begin{figure*}[!htp]
\includegraphics[width=6.5in]{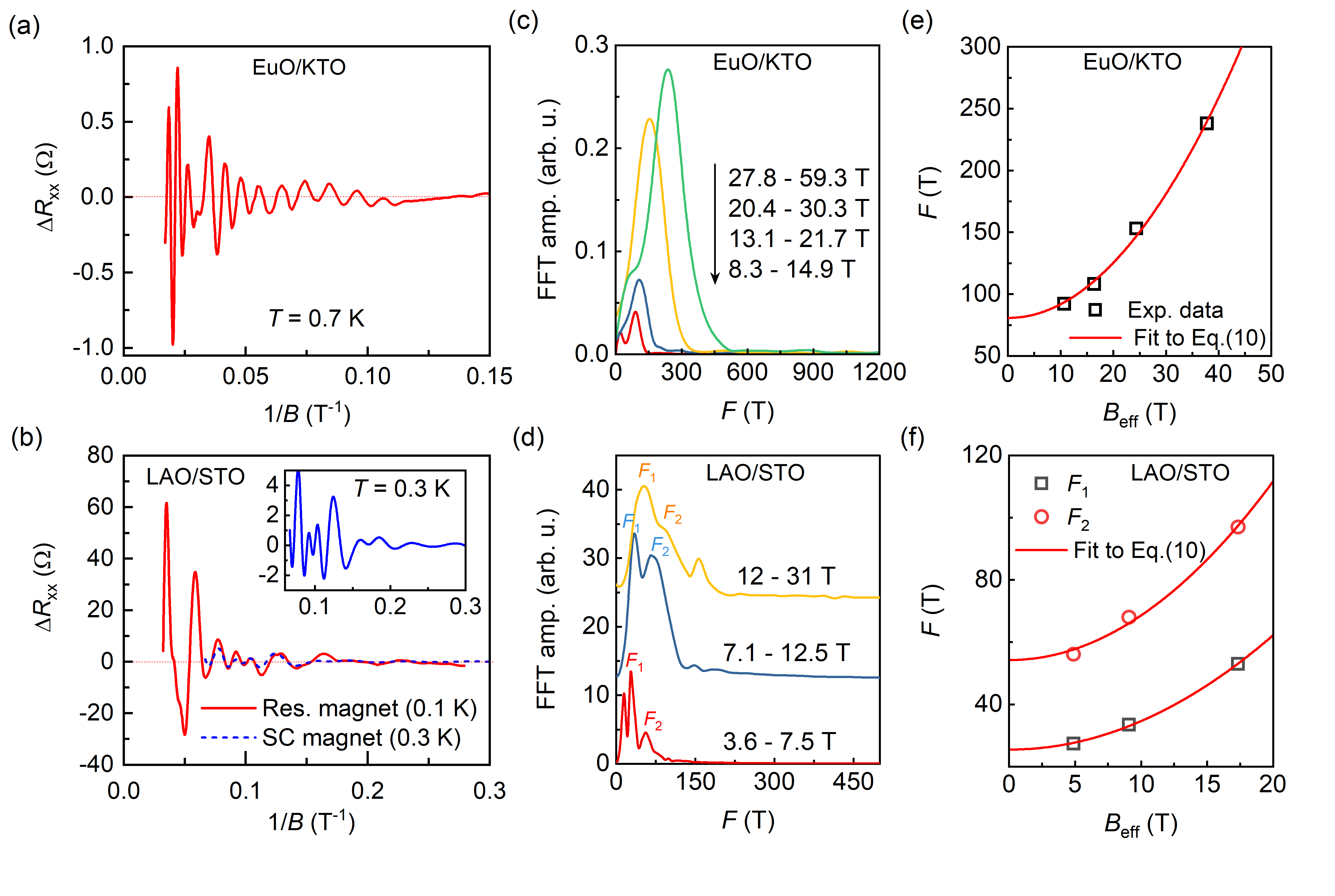}
\caption{\label{F2} Aperiodicity in quantum oscillations from 2D confined electrons at LAO/STO and EuO/KTO interfaces. Left column: $\Delta R_{xx}$ as a function of inverse magnetic field ($1/B$) for (a) EuO/KTO and (b) LAO/STO. The inset in (b) is the LAO/STO data measured in a superconducting magnet in field up to 14 T at 0.3 K. Middle column: fast Fourier transform (FFT) analysis in different windows of the $\Delta R_{xx}$($1/B$) data shown in left column. Right column: Frequency of oscillations estimated from the peak position in FFT analysis as a function of $B_{eff}$ for (e) EuO/KTO and (f) LAO/STO. In (e) and (f), symbols are the frequency estimated from the experimental data and solid lines are the fit to Eq. (10).}
\end{figure*}

Since the temperature dependence of quantum oscillations amplitude provides a means for determining the effective mass, we measure both systems at different temperatures. In particular, we measure EuO/KTO at various selected temperatures for two different field orientations $\theta = 0^{\circ}$ and $90^{\circ}$  and show the oscillations resistance in Fig. 3(a) and (b). It is to be noted that to improve the signal-to-noise ratio in pulsed magnetic fields, the measurements on EuO/KTO at different temperatures were performed using a higher frequency (256 kHz) of excitation. The higher frequency excitation did not modify the frequency and amplitude of quantum oscillations, as compared in Fig.A3(b) of Appendix A1. As expected, the oscillations amplitude progressively decreases with increasing temperature for $\theta=0^{\circ}$. We, however, noticed a nonmonotonic temperature dependence of the oscillations amplitude for $\theta = 90^{\circ}$, most likely due to imperfect subtraction of the smooth background from the raw data. Overall, the oscillations amplitude and frequency at $\theta = 0^{\circ}$ are larger than those at $\theta = 90^{\circ}$, and therefore, we assume that the oscillations from the carriers confined at the interface dominate at $\theta = 0^{\circ}$. 
We determine the cyclotron mass $m_c$ by fitting the temperature dependence of oscillations amplitude to the temperature damping factor in the Lifshitz-Kosevich (L-K) equation \cite{shoenberg2009magnetic} given below
\begin{equation}
R(T) = R_0 \frac {2{\pi}^2k_Bm_cT/\hbar eB}{sinh(2{\pi}^2k_Bm_cT/\hbar eB)} 
\end{equation}

For $\theta = 90^{\circ}$, we fit the maxima-to-minima difference to minimize the error in $m_c$ induced from the imperfect background subtraction. The $m_c$ values normalized with free electron mass $m_e$ are displayed in Fig. 3(e). At $\theta = 0^{\circ}$, $m_c$ = 0.56 $\pm$ 0.04 $m_e$ in moderate fields (10-14 T)  is comparable to the effective mass for heavy subbands (0.50 $m_e$) predicted theoretically \cite{PhysRevB.86.121107, PhysRevLett.108.117602} and confirmed with ARPES experiments \cite{PhysRevB.86.121107, PhysRevLett.108.117602} and SdH oscillations measurements \cite{harashima2013coexistence, rubi2021electronic, yan2022ionic} on KTO-2DES. 

Most interestingly, above 14 T, $m_c$ for $\theta = 0^{\circ}$ increases almost linearly, as depicted by a line, with increasing magnetic field strength. We, however, did not observe such a progressive field-dependent enhancement in $m_c$ values at $\theta = 90^{\circ}$. The average $m_c$ at $\theta = 90^{\circ}$ is 0.61$\pm$ 0.07 $m_e$, which corresponds well with the effective mass of the heavy band of bulk KTO \cite{PhysRevB.86.121107}. 

Next, we analyze the oscillations for LAO/STO measured at different temperatures in the range of 0.1 - 3.0 K (Fig. 4(a)). To ensure that the analysis is not primarily influenced by the superimposition of oscillations of multiple frequencies, we estimate $m_c$ for this system from the temperature dependence of FFT amplitude (Appendix A6). 
Very similar to EuO/KTO, the lowest value of $m_c$ (1.6 $\pm$ 0.1 $m_e$)  corresponds to the heavy subband of STO-2DES \cite{Delugas2011, rubi2020aperiodic}. Interestingly, $m_c$ for LAO/STO also increases with the magnetic field, bearing a resemblance with the data reported by Y. Xie \emph{et al.}\cite{xie2014quantum} in the moderate field range (4 - 15 T).

In conclusion, both interfaces exhibit a progressive enhancement of the cyclotron mass for $\theta = 0^{\circ}$ as the magnetic field intensifies. Since both the cyclotron mass $m_c$ ($= \frac{\hbar^2}{2\pi}\frac{\partial A_k}{\partial E}$) and the frequency of the quantum oscillations $F$ ($=\frac{\hbar}{2\pi e}A_k$) are related to the $k$-space area enclosed by the cyclotron orbit $A_k$, we next examine any eventual variation of the oscillations periodicity with magnetic field.

\subsection{Aperiodicity in quantum oscillations}

From the semiclassical theory of Landau quantization developed by Onsager and Lifshitz \cite{onsager1952, shoenberg2009magnetic}, the oscillations in magnetoresistivity are periodic in $1/B$. To examine the periodicity of oscillations in the 2DES at EuO/KTO and LAO/STO interfaces, we plot $\Delta R_{xx}$ at $\theta = 0^{\circ}$ as a function of the inverse magnetic field in Fig. 5(a) and (b), respectively. As a quality check of the oscillations in LAO/STO in a low-field regime (< 14 T), we also measure the same sample in a superconducting magnet at a temperature of 0.3 K and display this data in the inset as well as in the main panel of Fig. 5(b). While for $B$ > 6 T, the minima and maxima of the oscillations from these measurements perfectly overlie with the high-field measurement data, we acquired better-resolved oscillations in low-fields ($B$ < 6 T). As one can see, for both interfaces, the oscillations period decreases as the magnetic field increases. We perform the FFT analysis for both systems in a few selected field ranges to evaluate the magnetic field dependence of the oscillations frequencies. The FFT spectra of EuO/KTO (Fig. 5(c)) reveal one or two peaks in each field window and the dominant peak position moves to higher frequency with decreasing average inverse field of selected windows. The shoulder peak (on the left of dominant peak) noticed in two field windows (8.3 - 14.9 T and 27.8 - 59.3 T) are most likely from the 3D oscillations as the FFT of oscillations at $\theta = 90^{\circ}$ produces peaks at the same frequencies (see Appendix A4). 

 Unlike EuO/KTO, the LAO/STO interface exhibits at least two prominent peaks for each field window and these peaks shifts to higher frequency with increasing field. {In previous reports of high-field magnetotransport in LAO/STO \cite{MingYang2016, rubi2020aperiodic}, some coauthors of this manuscript revealed non $1/B$-periodic oscillations from a single band. However, the current data measured at ultralow temperatures (0.1 K) successfully resolve aperiodic oscillations originating from two different subbands. Additionally, the fully resolved oscillations at low temperatures discard the poor resolution of oscillations as one of the possible origins of aperiodic oscillations considered in the previous report \cite{rubi2020aperiodic}.}
In Fig. 5(e) and 5(f), we plot the estimated frequencies as a function of the effective field $B_{eff}$ for EuO/KTO and LAO/STO, respectively.  $B_{eff}$ defines as $\frac{1}{B_{eff}}=\frac{\frac{1}{B_{min}}+\frac{1}{B_{max}}}{2}$ depends on the size of the field range used in the FFT analysis.  It is worth mentioning that the FFT analysis of the data at  $\theta = 0^{\circ}$ in the full field range gives 7-8 peaks for both interfaces (Appendix A7, Fig A10) because of the progressive increase in oscillations frequency with field. Contrary to the observation at $\theta = 0^{\circ}$, the FFT analysis of the oscillations at $\theta = 90^{\circ}$ reveals only two frequencies (Appendix A4, Fig A6). 
In conclusion, the FFT analyses for the data at $\theta = 0^{\circ}$ reveal that the 2DESs at the studied interfaces exhibit a continuous increase in quantum oscillations frequency as the magnetic field strength rises, in line with the previous observation on LAO/STO interface \cite{fete2014large, MingYang2016, trier2016quantization, cheng2018shubnikov, rubi2020aperiodic}. 

To rule out the influence of 3D carriers and ensure the robustness of the key findings in our study of EuO/KTO, we measured a lower density (1.07 $\times$ 10$^{13}$ cm$^{-2}$) and higher mobility ($\sim$ 11,000 cm$^{2}$V$^{-1}$s$^{-1}$) KTO-2DES created at the \emph{a}-LAO/KTO. To enhance the mobility of the as-grown $a$-LAO/KTO samples, we control the oxygen vacancies at the interface using ionic-liquid gating. Details on the sample growth and the ionic-liquid gating treatment are given in Ref. \cite{yan2022ionic}. Notably, the density of electrons at the \emph{a}-LAO/KTO interface is comparable to that of LAO/STO. As displayed in Fig. 6, the \emph{a}-LAO/KTO interface demonstrates aperiodic oscillations and mass enhancement with magnetic field, resembling the results observed at the EuO/KTO and LAO/STO interfaces. Furthermore, the mass $\sim 0.52 m_e$ at low fields is the same as that for EuO/KTO. However, the field dependence of the mass differs between these systems. We attribute this difference to the varying spin-orbit coupling. Since the spin-orbit coupling is highly dependent on carrier density \cite{PhysRevLett.104.126803, nakamura2009}, the one-order-of-magnitude lower carrier density in the \emph{a}-LAO/KTO sample compared to the EuO/KTO results in the observed differences in the field dependence of mass.

\begin{figure}[!htp]
\includegraphics[width=3.4in]{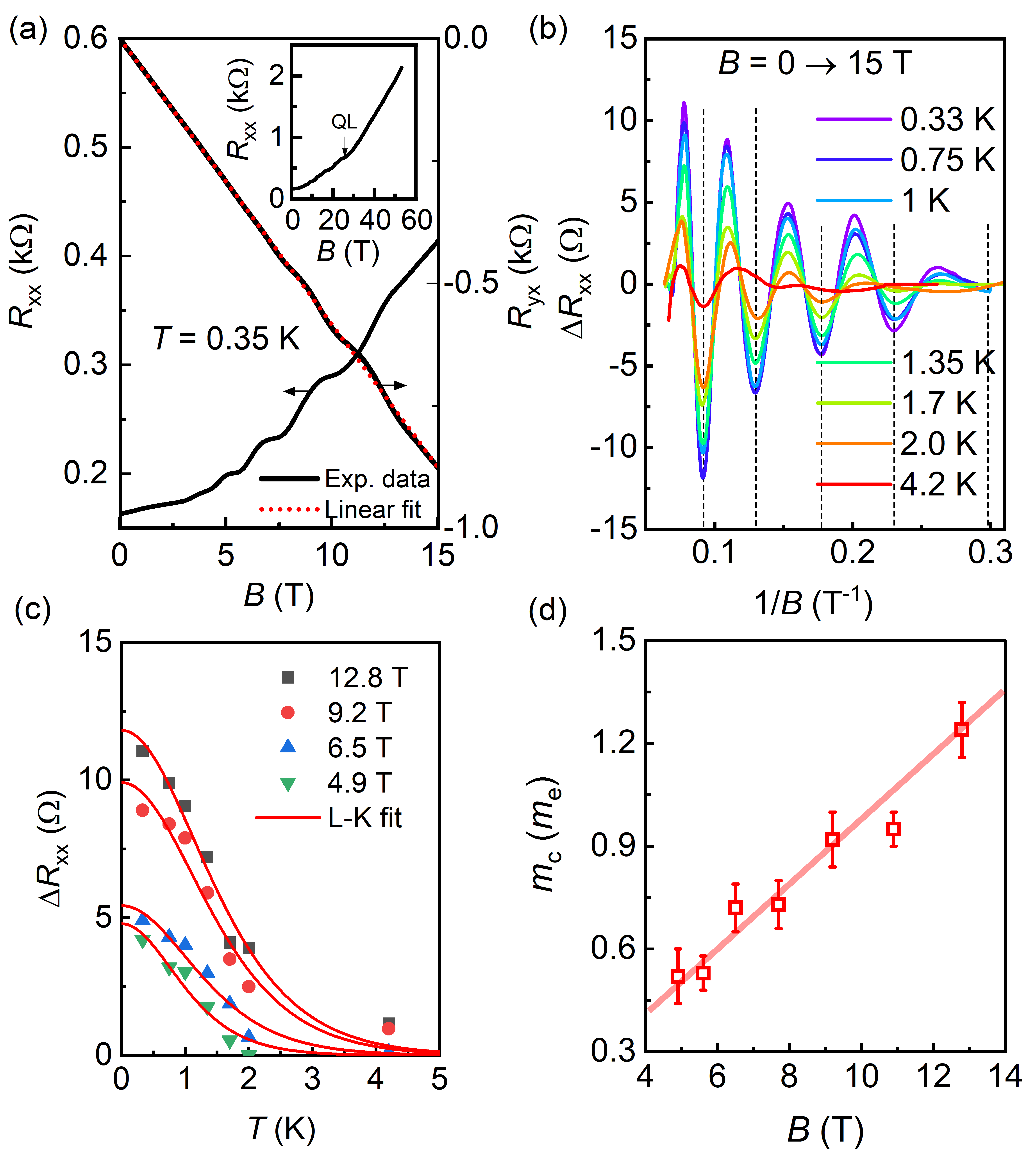}
\caption{\label{F2} {Aperiodic quantum oscillations and mass enhancement in \emph{a}-LAO/KTO. (a) Magnetic field dependence of longitudinal resistance $R_{xx}$ (left y-axis) and Hall resistance $R_{yx}$ (right y-axis) at $T$ = 0.35 K. The quantum oscillations are clearly visible in the raw data of $R_{xx}(B)$ and $R_{yx}(B)$. Inset shows the $R_{xx}$($B$) data measured up to the field of 55 T. No oscillations are observed above 28 T field, (b) Inverse field dependence of oscillating resistance at different temperatures, (c) Temperature dependence of oscillations amplitude at different extrema points fitted with the L-K formula, and (d) Cyclotron mass enhancement with magnetic fields. }}
\end{figure}

\section{Interpretation of unconventional findings from quantum oscillations}

The shared unconventional findings from the analysis of the quantum oscillations in 2DES at all three investigated interfaces are as follows: (1) The cyclotron mass estimated at low field values is comparable to the effective mass for heavy subbands. (2) Both the cyclotron mass and the oscillations frequency {in $1/B$} increase with the magnetic field. These observations appear to be universal, regardless of sample mobility and density, at least within the regime achievable in our samples.

The quantum oscillations resolved only from the heavy subbands can be attributed to the low carrier mobility in the light subbands. Despite a lighter effective mass, electrons in these subbands, mainly composed of $d_{xy}$ orbitals, exhibit reduced mobility due to their existence in the interface-adjacent planes (TiO$_2$ planes for STO and TaO$_2$ planes for KTO), which typically experience significant disorder (e.g., intermixed ions and dislocation) induced during the growth of the top oxide layers \cite{chen2016, rubi2020aperiodic, liu2021two}. 
Of particular interest are the mass enhancement and the large cyclotron mass observed at high magnetic fields. In the case of EuO/KTO, the cyclotron mass reaches approximately 1.8 $m_e$, while in LAO/STO, it reaches around 3.0 $m_e$. These values cannot be explained solely based on the predicted mass of electronic subbands \cite{Delugas2011, santander2012orbital} or magnetic breakdown \cite{xie2014quantum}. Furthermore, as explained in Appendix A8, the aperiodicity in oscillations can also not be explained by considering an internal magnetic field induced by magnetic moment at the interface. While the magnetic-field-induced change in density or chemical potential can reasonably explain the increase in oscillations frequency, the mass enhancement contradicts this scenario if the electronic bands follow a parabolic dispersion relation, for which $\frac{\partial A_k}{\partial E}$ is constant. Therefore, the magnetic-field-induced simultaneous change in frequency and cyclotron mass (\emph {i.e.} change in $A_k$ and $\frac{\partial A_k}{\partial E}$) implies a correction to the parabolic dispersion of the electronic bands. 

We consider a {linear correction to} parabolic dispersion to interpret the $B$ dependence of $A_k$ and $\frac{\partial A_k}{\partial E}$.
Combining parabolic and linear dispersion terms {with the Zeeman effect}, the Hamiltonian for a 2DES in a magnetic field perpendicular to its plane will be \cite{taskin2011berry, tisserond2017aperiodic, PhysRevB.109.115303} 
\begin{equation}
H = \frac{\Pi^2}{2m}+ v_F(\Pi_x\sigma_y + \Pi_y\sigma_x) - \frac{1}{2}g\mu_B B\sigma_z
\end{equation}
\\
where $\Pi_i = \hbar k_i+ eA_i$, $m$ is the density of states (DOS) mass, $v_F$ is the Fermi velocity, $\sigma_i$ are the Pauli matrices, $g$ is the Landé $g$-factor, and $\mu_B$ is the Bohr magneton. 
 
 The associated Landau levels for the Hamiltonian in Eq. (2) will be \cite{taskin2011berry}
\begin{equation}
 E_N = \hbar\omega_SN\pm\sqrt{(\hbar\omega_D)^2N+\left(\frac{\hbar\omega_S}{2} - \frac{g\mu_BB}{2}\right)^2}
 \end{equation}
 where $\omega_S = eB/m$, $\omega_D = \sqrt{2ev_F^2B/\hbar}$, and $N$ is the Landau level index.
 Taking $E_N = E_F$ and converting Eq. (3) as a quadratic equation for $N$, we get 
 
\begin{widetext}
 \begin{equation}
 (\hbar\omega_S)^2N^2 - [2\hbar \omega_S E_F + (\hbar \omega_D)^2]N + 
 E_F^2-\frac{1}{4}(\hbar \omega_S - g\mu_BB)^2 = 0
 \end{equation}

  and by solving Eq. (4) for $N$, we have 
  

\begin{equation}
N = \frac{\frac{m^2v_F^2}{e\hbar}+\frac{mE_F}{e\hbar}}{B} + \frac{\sqrt{\left(\frac{m^2v_F^2}{e\hbar}\right)^2+2mE_F\left(\frac{mv_F}{e\hbar}\right)^2+\frac{1}{4}\left(1-\frac{mg\mu_B}{e\hbar} \right)^2B^2}}{B} .
\end{equation}

\end{widetext}

\begin{table*}[!htp]
\centering
\begin{tabular*}{\textwidth}{c @{\extracolsep{\fill}} c c c c c c c c c c}
 \hline\hline
 \multirow{2}{*}{Interface} & \multirow{2}{*}{Frequency} & \multicolumn{2}{c}{Landau plot} & \multicolumn{2}{c}{FFT analysis} &  \multicolumn{2}{c}{Carrier density} & {C. mass (LF)} & {Fermi Velocity}\\
 &  & $F_0 (T)$ & $C$ &  $F_0(T)$ & $C$ & $n_{SdH}$ (cm$^{-2}$) & $n_{Hall}$ (cm$^{-2}$)& {$m_c$ ($m_e$)} & {$v_F$ (m/s)}\\
 \hline\hline
 EuO/KTO  & $F_1$ & 76.6 $\pm$ 1.2     & -0.134 &   80.6 $\pm$ 3.8 &   -0.111  & $3.9 \times 10^{12}$ &$2.2 \times 10^{14}$ & 0.56 $\pm$0.04 &  {$5.4 \times 10^{4}$}\\
 \hline
 {{\em{a}}-LAO/KTO} & $ {F_1}$ & {14.8 $\pm$ 0.5} & {-0.139}  &-  & - &         $ {7.1 \times 10^{11}}$ &$ {1.0 \times 10^{13}}$ &  {0.52 $\pm$0.08} &  {1.$8\times 10^{4}$}\\
  \hline
 \multirow{2}{*}{LAO/STO} & $F_1$ &   -  & -  &  25.4 $\pm$ 0.4 &   -0.092 &  \multirow{2}{*}{$3.8 \times 10^{12}$ }  & \multirow{2}{*}{$3.1 \times 10^{13}$ } &  \multirow{2}{*}{1.59$\pm$0.21} & \multirow{2}{*}{ {$0.4\times 10^{4}$}} \\
 & $F_2$ &- & -&   54.2 $\pm$ 2.2 &  -0.143& & & &\\
 \hline\hline
\end{tabular*}
\caption{{Estimated parameters from the analysis of SdH oscillations of three interfaces, EuO/KTO, {\em{a}}-LAO/KTO, and LAO/STO}. $F_0$ and $C$ are extracted from fitting the Landau plot in Fig. 7(a) and the magnetic field dependence of frequencies in Fig. 5(e) and (f). The $n_{SdH}$ calculated using the formula $n_{SdH} = 2e\Sigma f_i/h $ is compared with $n_{Hall}$ extracted from the linear fit to the Hall resistance $R_{yx}(B)$. {Fermi velocity $v_F$ is estimated from the fitting of $m_c(B)$ to Eq.(13) as shown in Fig. 7(b).}}
\end{table*}

Considering that the first two terms in the square root of Eq. (5) are larger {than} the last one owing to the heavy DOS mass in KTO- and STO-2DES, we perform Taylor's expansion of the square root, and get an approximate expression for the Landau level index $N$:  
  \begin{equation}
 N \approx \frac{F_0}{B} + C \times B + .......
 \end{equation} 
 
 where
 
 \begin{equation}
 F_0=\frac{m^2v_F^2}{e\hbar}\left(1+\frac{E_F}{mv_F^2}+\sqrt{1+2\frac{E_F}{mv_F^2}}\right)
 \end{equation}
 and
 \begin{equation}
 C = \frac{e\hbar}{8m^2v_F^2}\frac{\left(1-\frac{mg\mu_B}{e\hbar}\right)^2}{\sqrt{1+\frac{2E_F}{mv_F^2}}}
   \end{equation}

 {From Eq. (7) and (8), we get }
   \begin{equation}
 {4F_0C \approx \left(1-\frac{m^*g}{2}\right)^2}
 \end{equation}
 {where $m^* = m/m_e$. }

\begin{figure}[!b]
\includegraphics[width=3.4in]{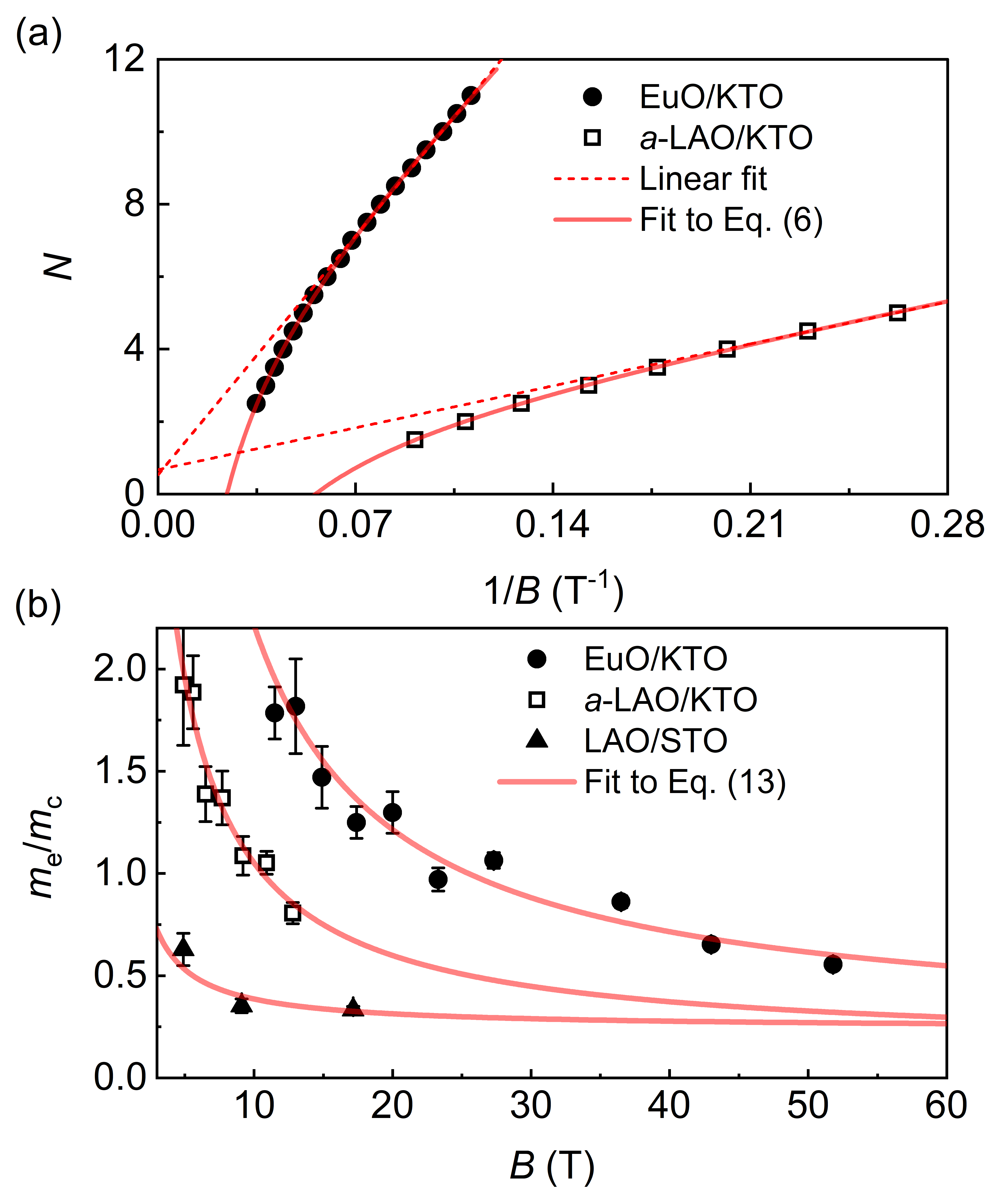}
\caption{\label{F2} (a) Landau plot (Landau level index versus $1/B$) for EuO/KTO {and \emph{a}-LAO/KTO} interfaces constructed assigning minima to integer and maxima to half-integer. Dashed lines are a linear fit to low-field data extrapolated to $B \rightarrow \infty$ ($1/B = 0$). The solid lines display a fit to the Eq. (6).  (b) Magnetic field dependence of the inverse cyclotron mass (symbols) with fit (solid line) to Eq. (13) for all three interfaces. {The impact of carrier density and effective mass on the field dependence of $1/m_c$ is distinctly visible.}}
\end{figure}
 
To be noted, Eq. (6) is the well-known Onsager’s relation \cite{onsager1952} with an additional term $C \times B$ that brings a deviation of the oscillations periodicity from 1/$B$ and leads to a non-linear Landau plot, $i. e.,$ a plot of Landau level index as a function of the inverse magnetic field. Constructing the Landau plot from the oscillations for two or more different frequencies (\emph{e.g.} EuO/KTO data for $B$ > 30 T in Fig. 5(a) and LAO/STO data in full-field range in Fig. 5(b)) is not feasible. We, therefore, display the Landau plot for EuO/KTO only for $B$ < 30 T and {\emph a-LAO/KTO for $B$ < 13 T} with reasonably good fit to Eq. (6) in Fig. 7(a) and list fitting parameters in Table 1. 
 
In order to determine a relationship between the oscillations frequency and magnetic field, we make a first order derivative of $N$ as a function of $1/B$, \emph{i.e.} 
 \begin{equation}
 F = \frac{\partial N}{\partial (1/B)} = F_0 - C \times B^2
 \end{equation}

Next, we apply this phenomenological model to the frequency extracted from the FFT analysis and show the best fit of the experimental data to Eq. (10) {for EuO/KTO and LAO/STO} in the Fig. 5(e) and (f). Interestingly, the fitting parameters $F_0$ and $C$ for EuO/KTO extracted from two different methods of analyzing quantum oscillations, namely the Landau plot and the FFT, are comparable (Table 1). Further, as displayed in Table 1 {and discussed in detail later}, the density calculated from the oscillations frequencies is smaller than the Hall carrier density for all interfaces, in line with previous reports \cite{fete2014large, MingYang2016, rubi2020aperiodic, harashima2013coexistence, yan2022ionic}. 
 
Next, to examine the field dependence of the cyclotron mass, we estimate the energy difference between two consecutive Landau levels, as given below 
 \begin{equation}
 E_{N+1}-E_N = \hbar\omega_c^*.
 \end{equation} 
 where $\omega_c^* = \frac{eB}{m_c^*}$ is the cyclotron frequency and $m_c^*$ is the effective cyclotron mass incorporating the effect of linear and parabolic dispersion of Hamiltonian in Eq. (2). It is important to note that the L-K formula in Eq. (1) is based on the effective mass theory with a parabolic dispersion. In the case of nonparabolic dispersion, the cyclotron mass extracted from the L-K analysis or cyclotron resonance naturally exhibits a dependence on energy and magnetic fields \cite{palik1961, rossner2006}.
 
  By substituting $E_N$ and $E_{N+1}$ in Eq. (10) and treating $\hbar\omega_D$ as a correction to $\hbar\omega_S$, we get an approximate expression for $\omega_c^*$:
 \begin{equation}
 \omega_c^* = \omega_S+\frac{\hbar \omega_D^2}{|\hbar \omega_S - g\mu_BB|}
 \end{equation}
 Using $\omega_c^* = \frac{eB}{m_c^*}$, $\omega_S = \frac{eB}{m}$, and $\omega_D = \sqrt{2ev_F^2B/\hbar}$, we obtain
 \begin{equation}
 \frac{1}{m_c^*} = \frac{1}{m}+\frac{mv_F^2}{|\hbar e - g\mu_B m|}\frac{1}{ B}
 \end{equation}
 
This expression for the cyclotron mass for $g$ = 0 is the same as derived directly using the cyclotron mass definition $m_c=\frac{\hbar^2}{2\pi}\frac{\partial A_k}{\partial E}$, where $A_k = \pi k^2$ and $E = \frac{\hbar^2k^2}{2m}+\hbar v_Fk$  (see Appendix A9). {From Eq. 13 and Eq. A5, the effective cyclotron mass predominantly depends on $v_F$, underscoring the significant impact of linear dispersion on mass enhancement. This assertion is reinforced by employing a pure linear model, as demonstrated in Appendix A11.}
By fitting the experimental $m_c$($B$) data for {all three interfaces to Eq. (13) in Fig. 7(b) and using $mg$ value estimated from Eq. (9), we calculate the Fermi velocity $v_F$ as listed in Table 1. As expected from the heavier mass for oxides-2DES, the estimated $v_F$} is one or two order of magnitude smaller than that for Dirac fermions in topological materials \cite{analytis2010, liu2014}.  

\begin{figure*}[!htp]
\includegraphics[width=6.5in]{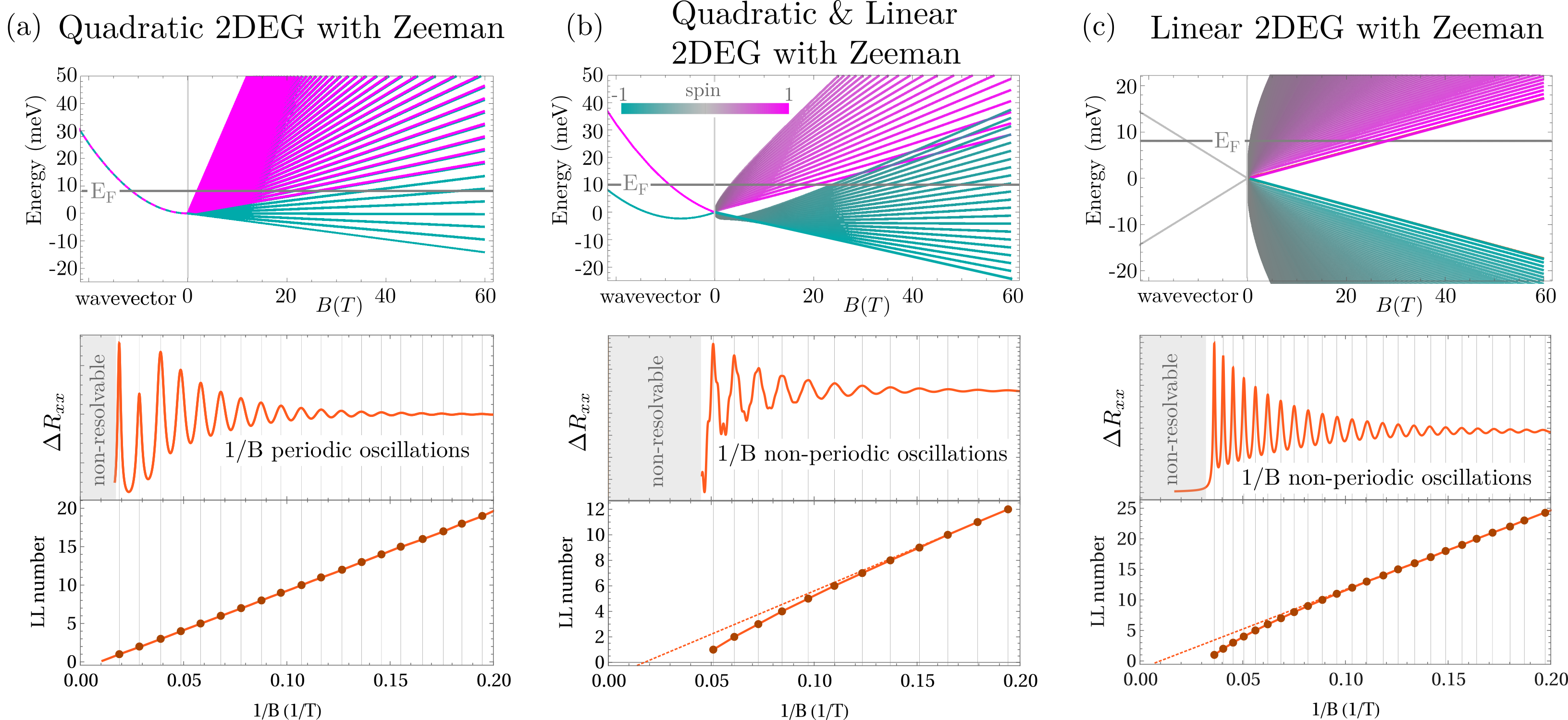}
\caption{\label{F7} {(a) Top panel: Plot of the electronic band structure and corresponding Landau Levels (LL) for a 2DEG of quadratic dispersion, with Zeeman interaction; middle panel: corresponding SdH-oscillations; bottom panel: LL index as a function of $1/B$. (b) Same as (a) for a 2DEG with both quadratic and linear dispersions, and Zeeman interaction. (c) For a linear 2DEG with Zeeman. The analytical calculations and the used parameters are summarised in Appendix A11. While the quadratic dispersion with Zeeman produces $1/B$-periodic oscillations and a linear Landau plot, the linear+quadratic and linear dispersions with Zeeman give rise to non $1/B$-periodic oscillations and non-linear Landau plot.}}
\end{figure*}

\section{Discussion and Conclusion}
{To understand the discrepancy between densities estimated from SdH oscillations and Hall resistance, we examine two primary scenarios (i) existence of 3D carriers, and (ii) unresolved oscillations from the low-energy subbands. For the first possibility, we calculate 3D-carriers density $n_{SdH}^{3D}$ for both EuO/KTO and LAO/STO interfaces using the oscillation frequencies for in-plane field orientation. EuO/KTO interface exhibits a higher  $n_{SdH}^{3D}$ ($8.7 \times 10^{18}$ cm$^{-3}$) compared to LAO/STO ($6.8 \times 10^{17}$ cm$^{-3}$). Assuming that the residual carriers ($n_{Hall} - n_{SdH}^{2D}$) are the 3D carriers, we estimate electron gas thicknesses of $\sim$ 39 $\mu$m for LAO/STO and $\sim$ 25 $\mu$m for EuO/KTO (see Appendix A4 for detailed calculations). However, such a significant extension of the electron gas seems implausible given its typical thickness of around 5 - 20 nm measured from direct techniques for similar carrier densities \cite{copie2009, ueno2011, chen2023}. Therefore, we infer that the disparity between Hall and SdH densities cannot solely be attributed to 3D carriers. Instead, it is more likely that the absence of a significant proportion of 2D carriers results from unresolved oscillations from the low-energy subbands, which, as mentioned earlier, exist in the interface-adjacent planes with significant disorders \cite{chen2016, rubi2020aperiodic, liu2021two}. To validate this, we compare Fermi pocket sizes estimated from oscillations with those reported from theoretical calculations. The largest Fermi pocket areas, derived from oscillation frequencies listed in Table-1, are 0.51 nm$^{-2}$ for LAO/STO and  1.2 nm$^{-2}$ for EuO/KTO, which are approximately $\sim$ 10-20 times smaller than the two outermost pockets for STO \cite{cancellieri2014, vaz2019} and KTO-2DEGs \cite{varotto2022}. This scenario supports the results of the two-band model fit to Hall resistance (Appendix A10) that reveals a density of low-mobility carriers one to two orders of magnitude larger than that of high-mobility carriers (SdH density). This postulation is also consistent with previous conclusions drawn from layer-resolved density of states and high-resolution transmission electron microscopy results \cite{rubi2020aperiodic}. Furthermore, as displayed in Fig. A13 of Appendix A11, a minor fraction (2 - 30$\%$, contingent on the $v_f$ value) of residual carriers can also be accounted for the underestimated value of $n_{2D}$ using the formula $n_{SdH} = 2e\Sigma f_i/h$ in the case of combined quadratic and linear dispersion.}

{Next, we explore the potential reasons for the non-conventional $E-k$ dispersion, \emph{i.e.} the combination of quadratic and linear terms as described in Eq. (2), close to the Fermi level. In the case of STO(001) and KTO(001) 2DES, this dispersion most likely arises from Rashba spin-orbit interaction and/or linear dispersion in the $\Gamma$-M direction combined with quadratic dispersions in $\Gamma$-X and $\Gamma$-Y directions within the first Brillion zone. The Rashba spin-orbit coupling in these systems emerges from the structure asymmetry of the effective quantum well confining the electrons and have been extensively reported \cite{PhysRevLett.108.117602, shanavas2014, lin2019, varotto2022, xu2024}. Additionally, in both systems,} the atomic spin-orbit interaction leads to partial avoidance of crossings between the light ($d_{xy}$) and heavy ($d_{xz}$ and $d_{yz}$) bands along $\Gamma$-M, resulting in an orbital dispersion reminiscent of Dirac dispersion \cite{vivek2017, vaz2019, johansson2021, kakkar2023}. Given that there are four such points in the Brillouin zone where a Dirac-like dispersion occurs, it is plausible that electrons orbiting within the electronic states reconstructed from the combination of $d_{xy}$ and $d_{xz}$/$d_{yz}$ will encounter an unusual summation of linear and parabolic dispersion. 
{The low-field mass and the Hall density values strongly suggest that the Fermi level in the studied systems lies above the third subband, where a linear dispersion along $\Gamma$-M has been predicted \cite{vaz2019, johansson2021, kakkar2023}.}

{To further validate the aperiodic oscillations from the Hamiltonian in Eq. (2), we performed theoretical SdH simulation. As displayed in Fig. 8 and Appendix A11, the non $1/B$-periodic oscillations arise naturally in a 2DEG exhibiting a dispersion combined with linear and quadratic terms. More precisely, while a pure quadratic dispersion with Zeeman interaction does not yield non $1/B$-periodic SdH oscillations, a linear dispersion with Zeeman does (as shown analytically in Appendix A11, Eq. A17). Given the characteristic of complex oxides 2DESs, such as (1) giant Rashba coupling $\sim$ 150 - 400 meV.$\AA$ \cite{lin2019, varotto2022, xu2024} and a linear dispersion along $\Gamma$-M direction, and (2) a large effective mass $m^* \sim 0.5-1.6 m_e$  due to the $d$-orbitals, the SdH oscillations arising from this model will be dominated by the linear and Zeeman terms, thus presenting non-periodic $1/B$ oscillations. This is in stark contrast to III-V semiconductors 2DEGs, which exhibit effective masses of $\sim 0.05 m_e$ and Rashba coupling of $\sim$ 100 meV.\AA \cite{PhysRevResearch.4.013039, candidoPRR2023, PhysRevB.109.115303}, thus generating SdH-oscillations periodic in $1/B$.} In more conventional III-V semiconductors, for instance InAs and InSb, the electron or hole systems experience non-parabolic bands (with higher-order corrections in $k$) due to the avoided anticrossings of light and heavy subbands \cite{palik1961, winkler2003, rossner2006}. {Therefore, to investigate whether the inclusion of non-quadratic $k$ terms, such as $k^4$, can account for aperiodicity in oscillations, we conducted a simulation of SdH oscillations using the Hamiltonian $H = \frac{\hbar^2k^2}{2m^*}+ \beta k^4 + \frac{1}{2}g\mu_B B\sigma_z$. The findings, detailed in Appendix A11, demonstrate that under realistic parameter values, the SdH oscillations exhibit perfect periodicity in $1/B$.}


In summary, to gain a deeper understanding of the electronic band structure of the 2DES based on {$d$ orbitals}, we conducted a thorough investigation of quantum oscillations in magnetoresistance of LAO/STO, EuO/KTO, and {\emph{a}-LAO/KTO} interfaces in high magnetic fields. By analyzing the observed oscillations at various tilt angles, we identified that {these} interfaces exhibit electron confinement in the two-dimensional plane at the interface, while a portion of carriers extends deep into the STO and KTO. Remarkably, for all interfaces, the oscillations originating from the 2D confined electrons display an increased frequency {in $1/B$} and cyclotron mass with increasing magnetic field strength. To explain these {universal and} peculiar findings, we have proposed a model involving a combination of linear and parabolic dispersion relations {with Zeeman interaction. Theoretically simulated SdH oscillations support the presence of both type of dispersions,} offering a reasonable explanation for the experimental observations. These results suggest the existence of non-trivial electronic states, potentially linked to the {Rashba spin-orbit interaction and/or the linear dispersion in $\Gamma$-M due to atomic spin-orbit interaction. Our experimental findings and versatile model hold promise for comprehending similar observations in other materials exhibiting non-trivial electronic states. This includes topological materials such as Dirac and Weyl semimetals for which strong spin-orbit interactions are a prerequisite.}


\section*{Acknowledgements}

We acknowledge support from the National High Magnetic Field Laboratory, supported by the National Science Foundation through NSF/DMR-1644779 and the state of Florida. K.R., M.K.C. and N. H. and pulsed field measurements were supported by the US Department of Energy "Science of 100 Tesla" BES program. We acknowledge the support of HFML-RU/FOM, member of the European Magnetic Field Laboratory. M.K.C. acknowledges support from NSF IR/D program while serving at the National Science Foundation. Any opinion, findings, and conclusions or recommendations expressed in these materials are those of the author(s) and do not necessarily reflect the views of the National Science Foundation. 

\section*{Appendices}

\renewcommand{\thesubsection}{A\arabic{subsection}}

\setcounter{subsection}{0}

\renewcommand{\thefigure}{A\arabic{figure}}

\setcounter{figure}{0}

\renewcommand{\theequation}{A\arabic{equation}}

\setcounter{equation}{0}

\subsection{{High-field} magnetotransport details for EuO/KTO}

\begin{figure}[!htp]
\includegraphics[width=2.8in]{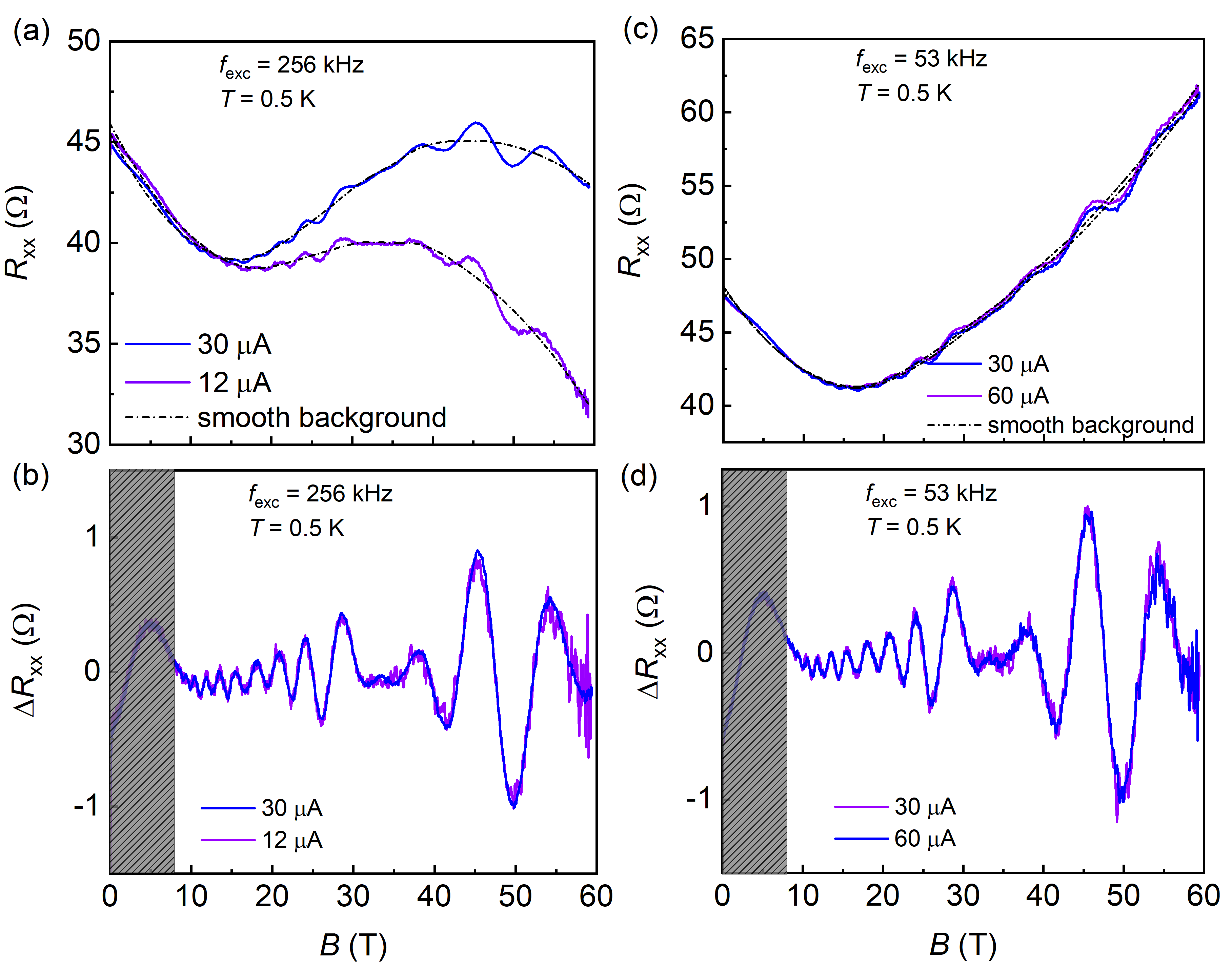}
\caption{\label{F} {Longitudinal resistance $R_{xx}$ measured by applying different excitation currents. (a) 12 $\mu$A and 30 $\mu$A at $f_{exc}$ = 256 kHz and (c) 30 $\mu$A and 60 $\mu$A at $f_{exc}$ = 53 kHz. Comparison of the oscillating resistance for different currents at (b) 256 kHz and (d) 53 KHz frequencies. The data at 256 kHz frequency shows a strong current-dependent magnetoresistance, however, the amplitude and period of the oscillations remain unaffected by the current, suggesting no Joule heating of the sample.}}
\end{figure}

\begin{figure}[!htp]
\includegraphics[width=3.0in]{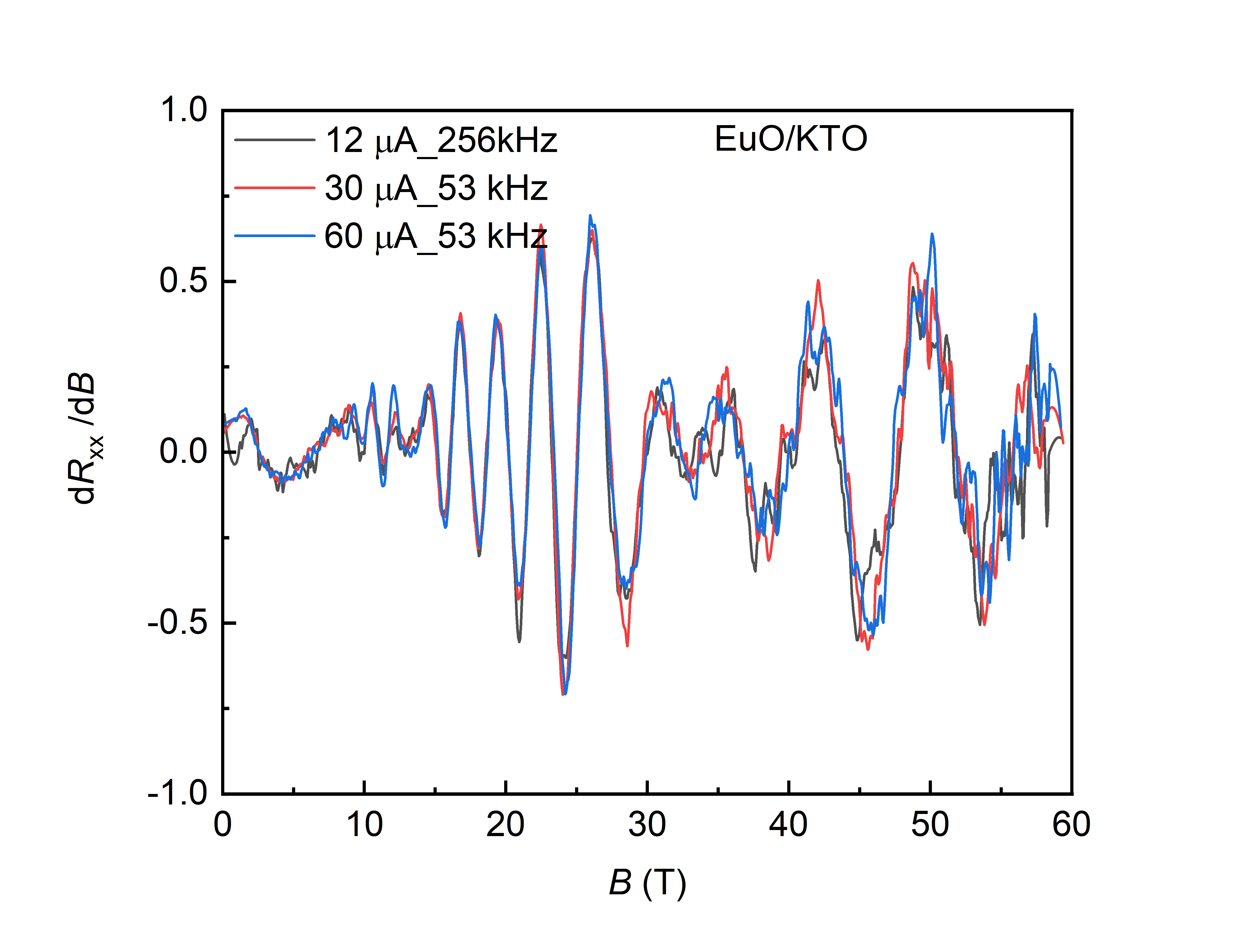}
\caption{\label{F} {First order derivative of $R_{xx}(B)$ for different excitation currents and frequencies. The oscillations amplitude and periodicity do not show any progressive shift with increasing current or frequency, and therefore, further signify absence of Joule heating in the sample.}}
\end{figure}

{To examine any apparent heating of the samples caused by the excitation current, we performed measurements applying different currents at the same temperature of 0.5 K. We compare the raw data of $R_{xx}$($B$) collected at two different currents in Fig. A1 (a) and (c) for two frequencies 256 kHz and 53 kHz, respectively. As one can see in Fig. A1(b), (d) and Fig A2, the higher excitation current $\sim$ 30 $\mu$A does not influence the amplitude of the oscillations that are highly susceptible to temperature.}

\begin{figure}[!htp]
\includegraphics[width=3.0in]{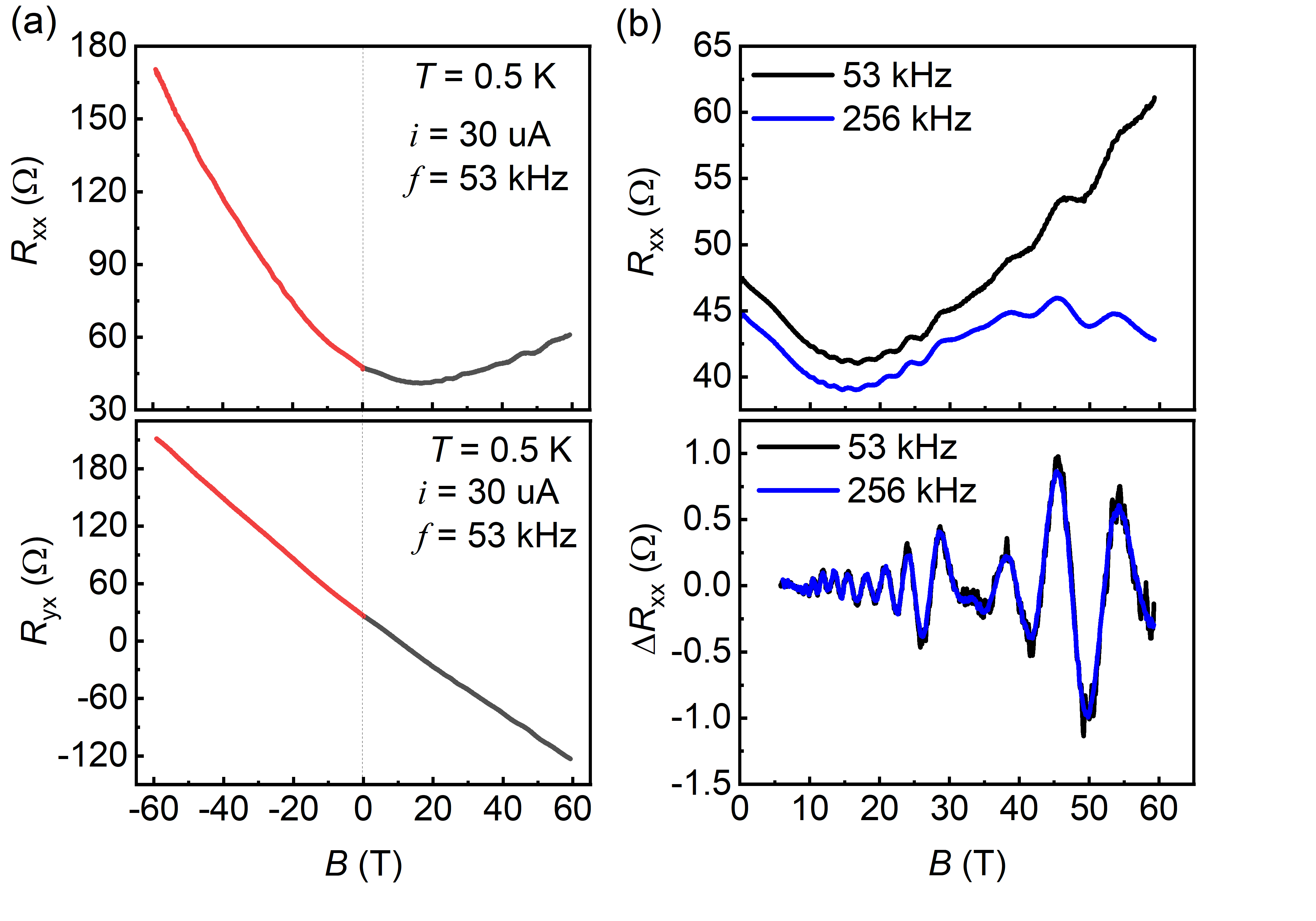}
\caption{\label{F} (a) Longitudinal resistance $R_{xx}$ (top) and Hall resistances $R_{yx}$ (bottom) measured in both field directions (up and down). (b) $R_{xx}$ (top) and $\Delta R_{xx}$ (bottom) for two different frequencies of excitation. The resistance absolute value and the magnetoresistance in high magnetic fields (> 30 T) is frequency dependent. The signal-to-noise ratio improves using high frequency excitation in the high pulsed magnetic field (pulse time $\sim$ 80 ms) and the oscillations amplitude remains unchanged.}
\end{figure}

In order to check the data symmetry in up and down fields for the unpatterned EuO/KTO sample, we measured transport on this sample in both field directions at the lowest possible temperature $T$ = 0.7 K and in the field perpendicular to the interface. The $R_{xx}$ and $R_{yx}$ (Fig. A3 (a)) both show asymmetry in the field. Despite the different magnitude of $R_{xx}$ and $R_{yx}$, we did not see a noticeable shift in the position or amplitude of oscillations. We used the asymmterized data (Fig. 1(b)), as described in the main text, to determine the carrier density and the mobility. Furthermore, since the high frequency of excitation improves the signal-to-noise ratio of the data measured in the pulsed magnetic field, we check its implication on the amplitude and frequency of oscillations. As displayed in Fig. A3 (b), we do not observe any noticeable change in the oscillations pattern except that the oscillations quality improves by increasing the frequency of excitation.


\subsection{{Sheet resistance of EuO/KTO interface measured using the Van der Pauw method}} 
\begin{figure}[!htp]
\includegraphics[width=3.0in]{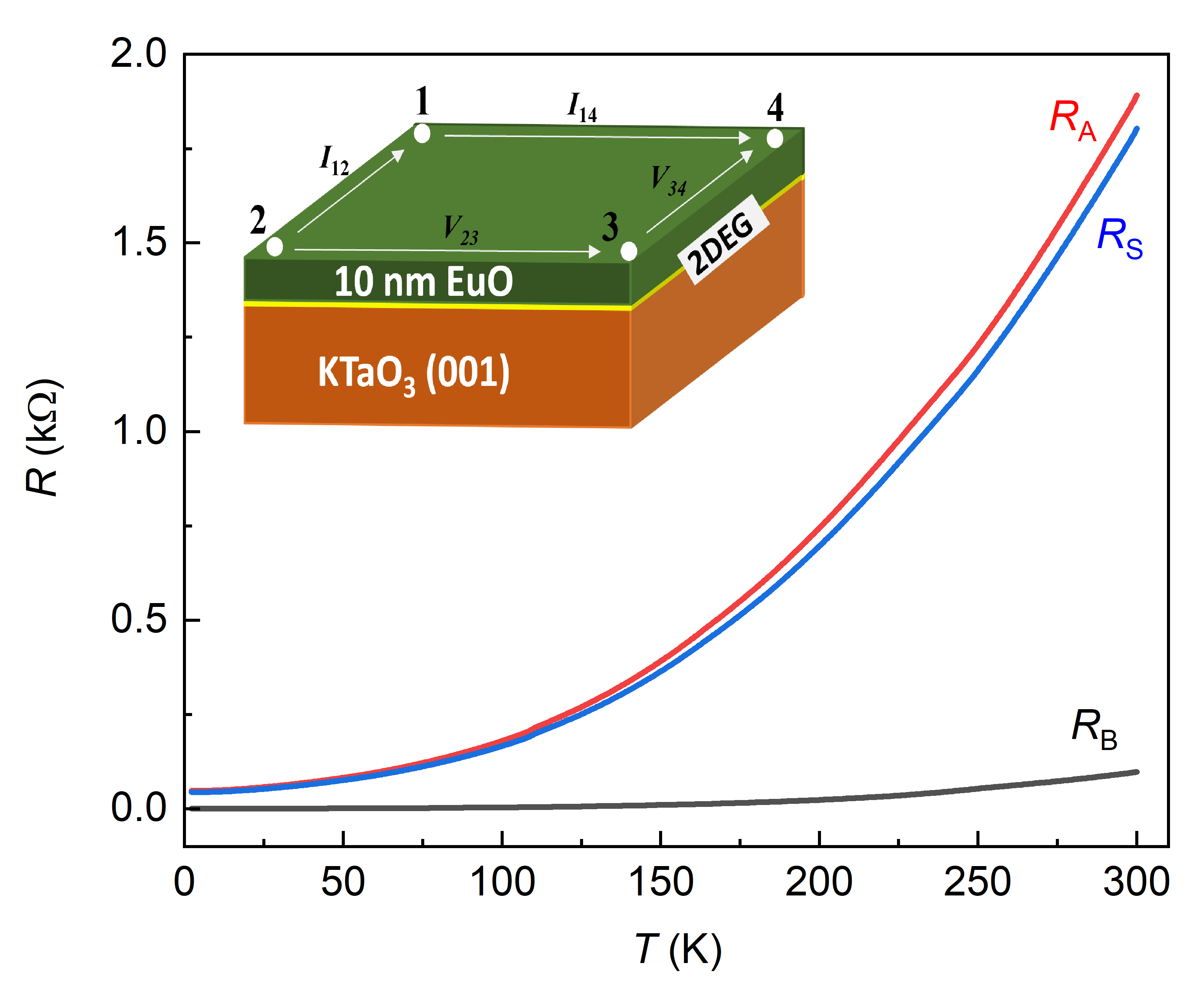}
\caption{\label{F2} {Temperature dependence of the longitudinal resistance of EuO/KTO measured in two different contact configurations and the calculated sheet resistance. }.}
\end{figure}

{To reduce errors in the mobility calculation of unpatterned EuO/KTO interface, we performed zero-field resistance measurements using the Van der Pauw method. We made contacts at the four corners numbered from 1 to 4 in a counter-clock-wise order as depicted in the inset of Fig. A4. The current is applied along one edge of the sample and the voltage is measured across the opposite edge. For example, the current $I_{14}$ is from contact 1 to 4 and the voltage $V_{23}$ is the voltage difference between contacts 2 and 3. The resistance is calculated as $R_A=\frac{V_{23}}{I_{14}}$ and $R_B=\frac{V_{34}}{I_{12}}$. We estimate the sheet resistance $R_S$ using an analytical method reported in \cite{oliveira2020} and a formula given below.}
\begin{equation}
{R_S=\frac{\pi}{ln2}\left(\frac{R_A+R_B}{2}\right)f\left(\frac{R_B}{R_A}\right)}
\end{equation}

{The estimated $R_S$ is displayed in Fig. A4 along with measured $R_A$ and $R_B$. The value of the $R_S$ 44.3 $\Omega / \square$ gives the mobility of $\sim$ 650 cm$^2$V$^{-1}$s$^{-1}$.}

\subsection{Out-of-plane and in-plane magnetoresistance}

\begin{figure}[!htp]
\includegraphics[width=3.0in]{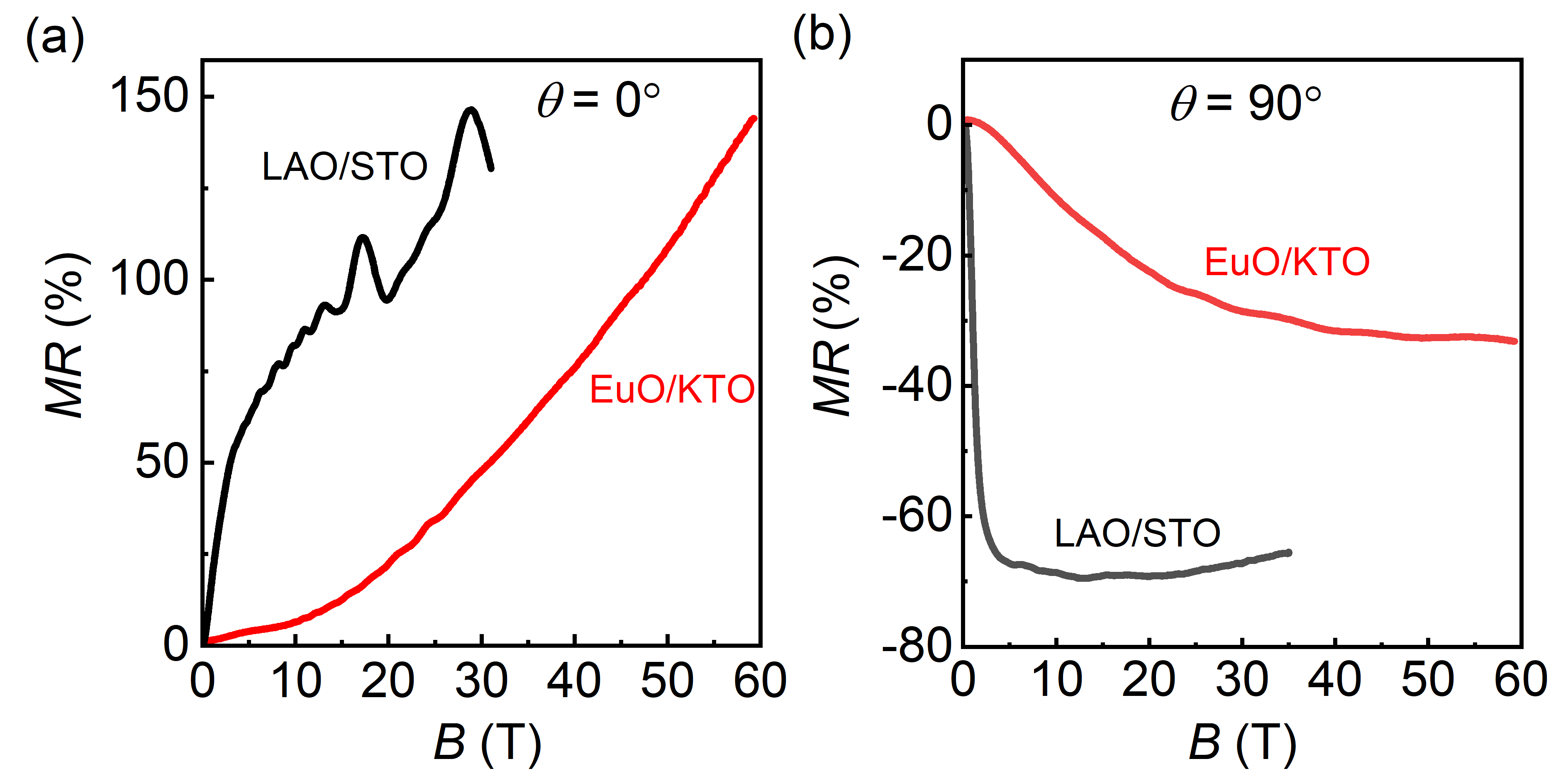}
\caption{\label{F2} A comparison of the magnetoresistance calculated as $MR = \frac{R(B)-R(B=0T)}{R(B=0T)}\times100$ for EuO/KTO and LAO/STO in (a) perpendicular-field orientation $\theta = 0^\circ$ and (b) parallel-field orientation $\theta = 90^\circ$. We observed a positive MR for $\theta = 0^\circ$ and negative MR for $\theta = 90^\circ$. The positive MR is almost linear in $B$ in high-field regime for both interfaces.}
\end{figure}

\subsection{FFT analysis of quantum oscillations at angles close to $\theta = 90^\circ$ {and 3D carrier density}}
To further verify that the position of SdH oscillations does not move by varying the angles in the vicinity of $\theta = 90^\circ$, we performed FFT analysis of the $\Delta R_{xx}$($1/B$) at a few angles. As shown in Fig A6, the position of the prominent peaks for both interfaces does not move with the angle. 
{We estimate the density of 3D carriers using the oscillations frequencies from FFT analysis and the formula }
\begin{equation}
{n_{SdH}^{3D} = \frac{8}{3\sqrt \pi}\left(\frac{ef}{h}\right)^{3/2}.}
\end{equation}

\begin{figure}[!htp]
\includegraphics[width=3.0in]{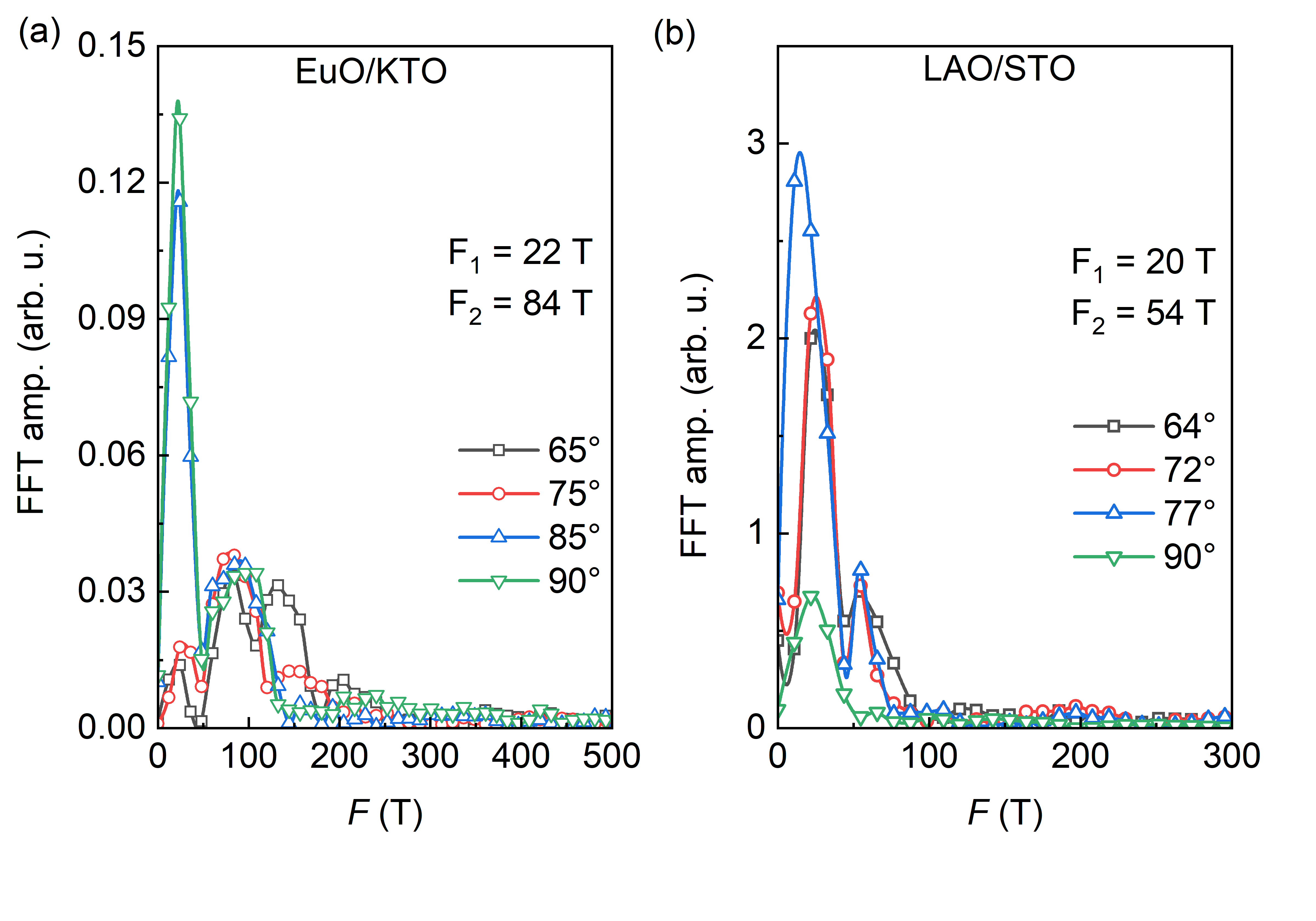}
\caption{\label{F2} Tilt-angle dependence of FFT spectra close to in-plane magnetic field, $\theta = 90^\circ$ for (a) EuO/KTO and (b) LAO/STO. }
\end{figure}
As expected from the Hall density (Fig. 1c and d), the value of $n_{SdH}^{3D}$ for EuO/KTO ($8.7 \times 10^{18}$ cm$^{-3}$) is higher compared to that for LAO/STO ($6.8 \times 10^{17}$ cm$^{-3}$). Next, assuming that the residual 2D carrier density ($n_{2D}=n_{Hall} - n_{SdH}^{2D}$) is equal to  $n_{SdH}^{3d}$, we estimate the electron gas thickness $\left(t = \frac{n_{2D}}{n_{3D}}\right)$ of $\sim 25 \mu$m for EuO/KTO and $\sim$ 39 $\mu$m for LAO/STO. Comparing the electrons gas thickness of a few nm reported in the literature for similar carrier density, we conclude that the discrepancy between Hall and SdH density can not be solely explained by the fact of coexistence of 2D and 3D carriers.

\subsection{2D confinement of electrons}
\begin{figure}[!b]
\includegraphics[width=3.5in]{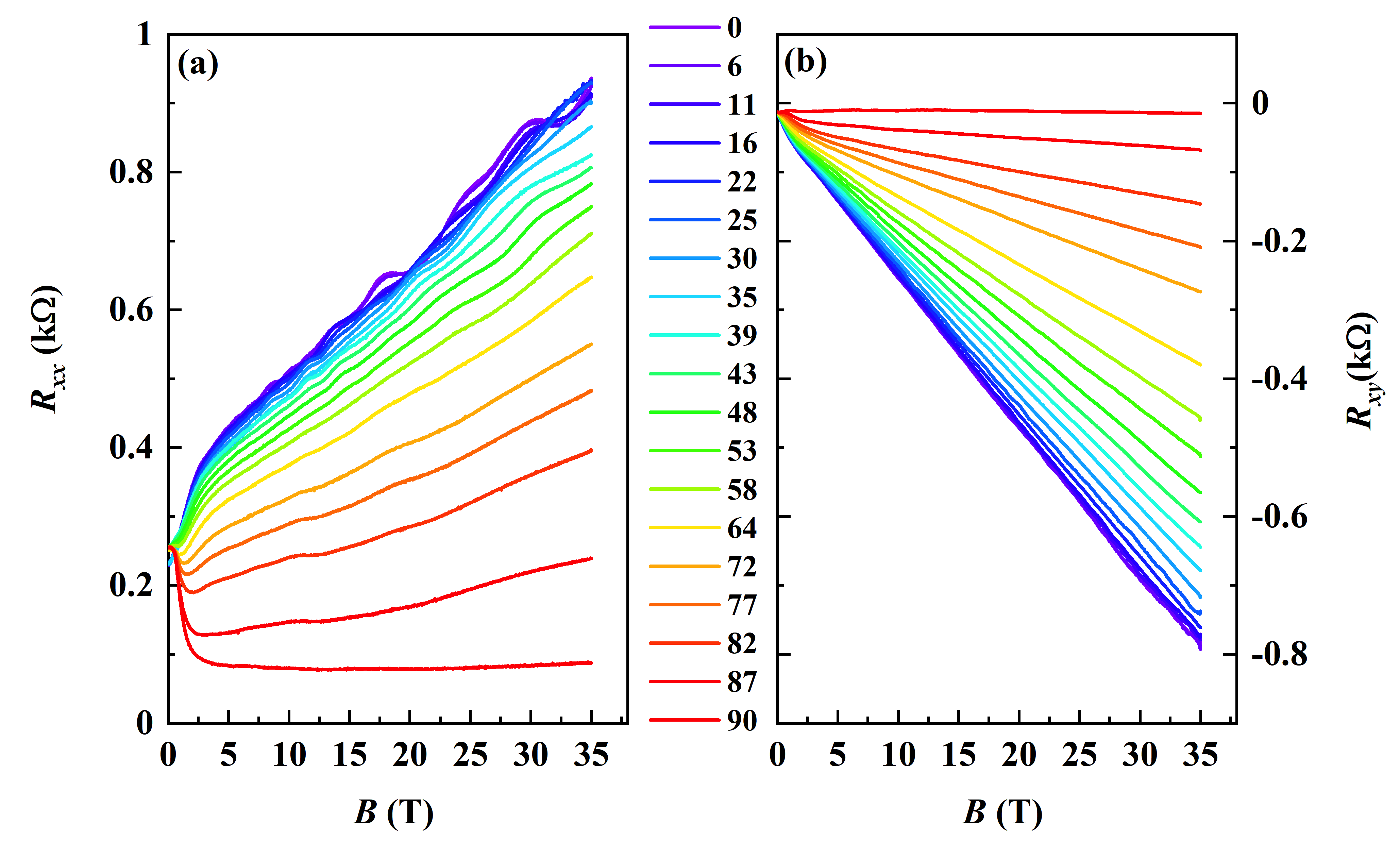}
\caption{\label{F2} Field dependence of (a) $R_{xx}(B)$ and (b) $R_{xy}(B)$ for different angles ranging from $\theta = 0^\circ$ to $\theta = 90^\circ$ for LAO/STO interface. The positive MR decays and switches to negative as magnetic field orientation moves from $\theta = 0^\circ$ to $90^\circ$. The Hall resistance also decreases progressively with changing field orientation and becomes almost zero at $\theta = 90^\circ$, indicating the 2D confinement of majority of electrons.}
\end{figure}

To further conclude that the majority of electrons are confined in 2D at the interface, we show an angular dependence of $R_{xx}(B)$ and $R_{xy}(B)$ for LAO/STO in Fig. A7 and the quantum oscillations in $R_{xy}(B)$ for all three studied systems in Fig. A8. As the magnetic field orientation varies from $\theta = 0^\circ$ to $90^\circ$, a progressive decrease in both $R_{xx}(B)$ and $R_{xy}(B)$ is observed, and most importantly, the $R_{xy}(B)$ becomes almost zero at $\theta = 90^\circ$ suggesting a 2D confinement of carriers. The existence of quantum oscillations in the Hall resistance of all three studied systems (Fig. A8) further confirms the 2D confinement of electrons. 

\begin{figure}[!htp]
\includegraphics[width=2.3in]{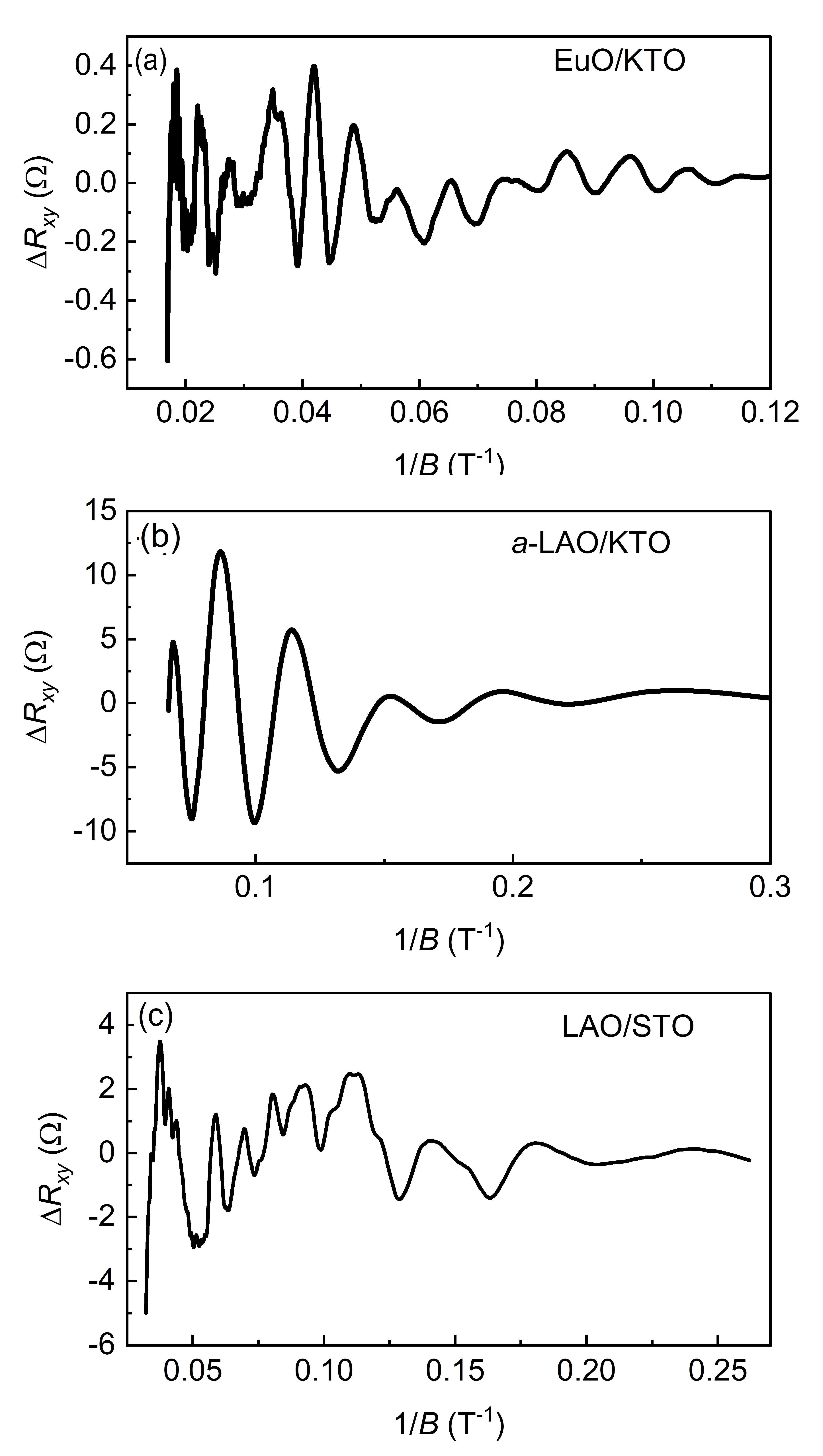}
\caption{\label{F2} Quantum oscillations in Hall resistance $R_{xy}$ obtained by subtracting a linear fit to the $R_{xy}(B)$ of EuO/KTO (Fig. 1c), a-LAO/KTO (Fig. 6a), and LAO/STO (Fig. 1d). These oscillations show a phase shift of $\pi/2$ with oscillations in $R_{xx}$  as expected from the relation $R_{xx}\propto \frac{dR_{xy}}{dB} \times B$ for 2D systems.}
\end{figure}


\subsection{L-K fit to FFT amplitude for LAO/STO}
\begin{figure}[!htp]
\includegraphics[width=3.0in]{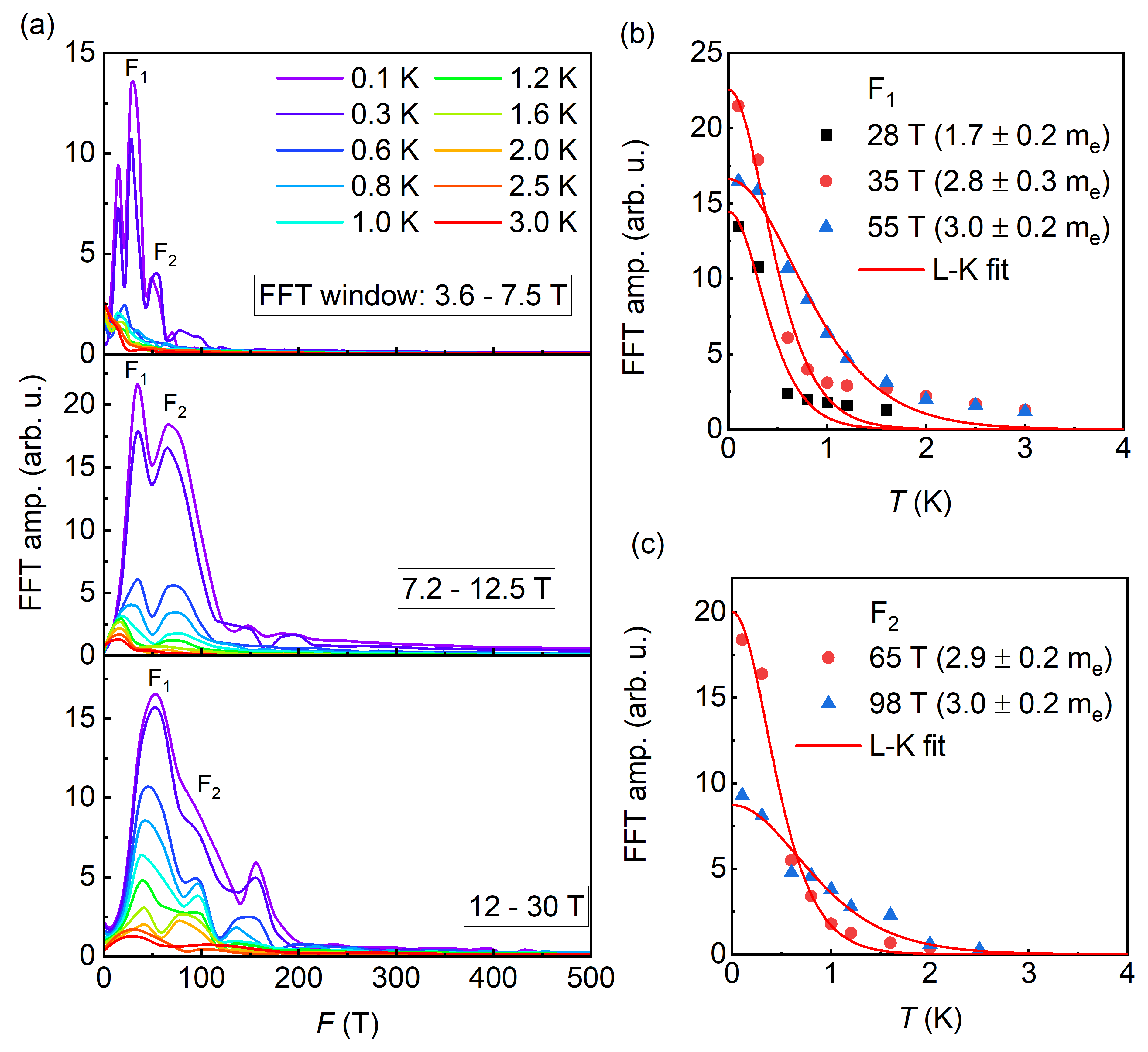}
\caption{\label{F2} Cyclotron mass calculation for LAO/STO from the FFT spectra in different windows.}
\end{figure}

In order to evaluate the cyclotron mass $m_c$ from the FFT spectra in Fig. A9(a), we fit the temperature dependence of FFT amplitude with L-K equation given below:
\begin{equation}
X(T) = X_0 \frac {2{\pi}^2k_Bm_cT/\hbar eB_{eff}}{sinh(2{\pi}^2k_Bm_cT/\hbar eB_{eff})} 
\end{equation}
where $\frac{1}{B_{eff}}=\frac{\frac{1}{B_{min}}+\frac{1}{B_{max}}}{2}$. The calculated $m_c$ values are displayed in Fig. A9(b) and (c) for frequencies $F_1$ and $F_2$, respectively. 


\subsection{FFT analysis of quantum oscillations in full-field range}
\begin{figure}[!htp]
\includegraphics[width=3.2in]{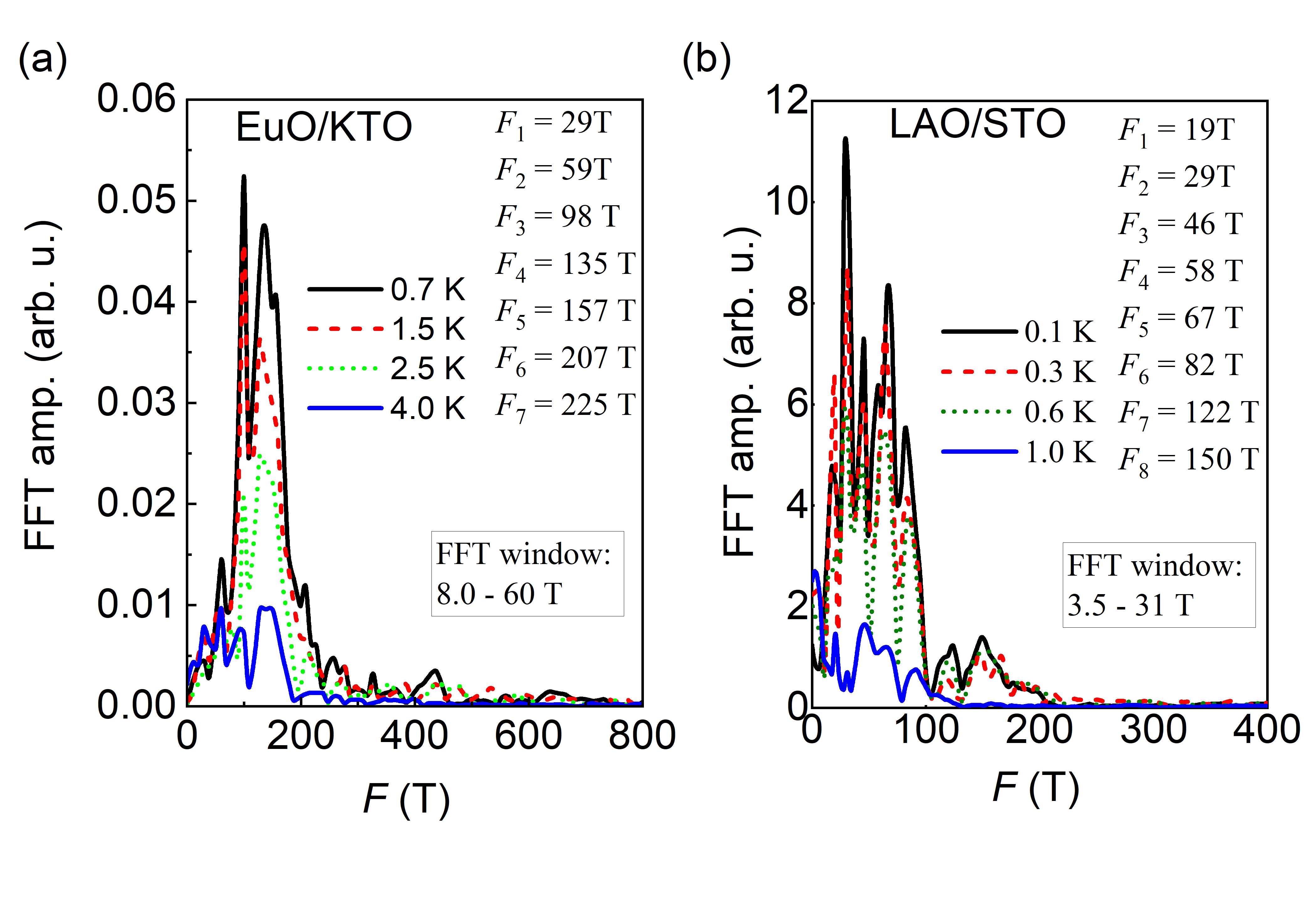}
\caption{\label{F2} FFT analysis of $\Delta R_{xx}$($1/B$) measured at $\theta = 0^\circ$ and at a few selected temperatures for (a) EuO/KTO and (b) LAO/STO.}
\end{figure}
\subsection{Linearization of Landau plot}
\begin{figure}[!htp]
\includegraphics[width=3.0in]{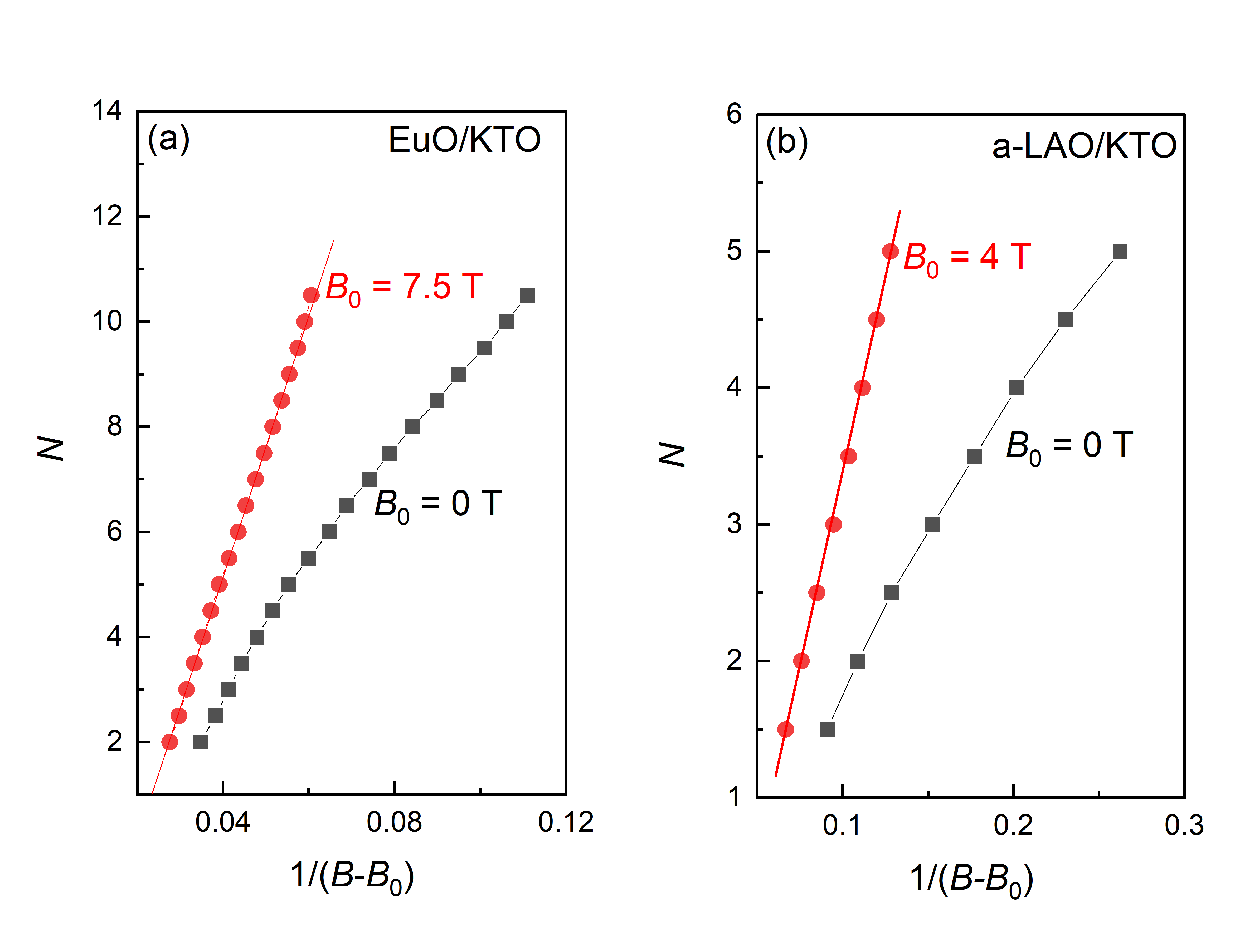}
\caption{\label{F2} Reconstructed Landau plot where Landau level index are plotted against $1/(B-B_0)$ for (a) EuO/KTO and (b) a-LAO/KTO. $B_0$ is the internal magnetic field induced by magnetization.}
\end{figure}
The aperiodicity in quantum oscillations could be due to the internal magnetic field induced by magnetization, like the case of ferromagnetic metal. To verify this scenario, we linearize the Landau plot by plotting $N$ as a function of $1/(B-B_0)$, where $1/B_0$ is internal magnetic field. As shown in Fig. A11, the Landau plot can be linearlized by considering extremely large values of $B_0$ (e.g., 7.5 T for EuO/KTO and 4 T for a-LAO/KTO). For reference, the $B_0$ estimated from quantum oscillations in Fe is about 2 T \cite{anderson1963}. Since the 2DES at the oxide interfaces cannot be more magnetic than Fe in any scenario, we exclude the possibility of internal magnetic field modifying the periodicity of oscillations in the studied system.

\subsection{Cyclotron mass in the case of combined linear and parabolic dispersion}
Combining the linear and parabolic dispersion term, we have $E = \frac{\hbar^2k^2}{2m}+\hbar v_Fk$.  The area of the Fermi surface in k space can be given as $A_k = \pi k^2$. Substituting $E$ and $A_k$ values in the cyclotron mass formula, we get
\begin{equation}
\frac{1}{m_c}=\frac{2\pi}{\hbar^2}\frac{1}{\partial A_k/\partial E} = \frac{1}{m}+\frac{v_F}{\hbar k}
\end{equation}
As we know, $\omega_c = \frac{v}{r} = \frac{eB}{m}$. Converting $r$ into reciprocal space, we have $k \approx \frac{eB}{mv_F}$. Substituting $k$ value into Eq. (A2), we get 
\begin{equation}
\frac{1}{m_c} \approx \frac{1}{m}+\frac{mv_F^2}{e \hbar}\frac{1}{B}
\end{equation}


\subsection{{Two-band model fit to the Hall resistance}}

\begin{figure}[!b]
\includegraphics[width=3.4in]{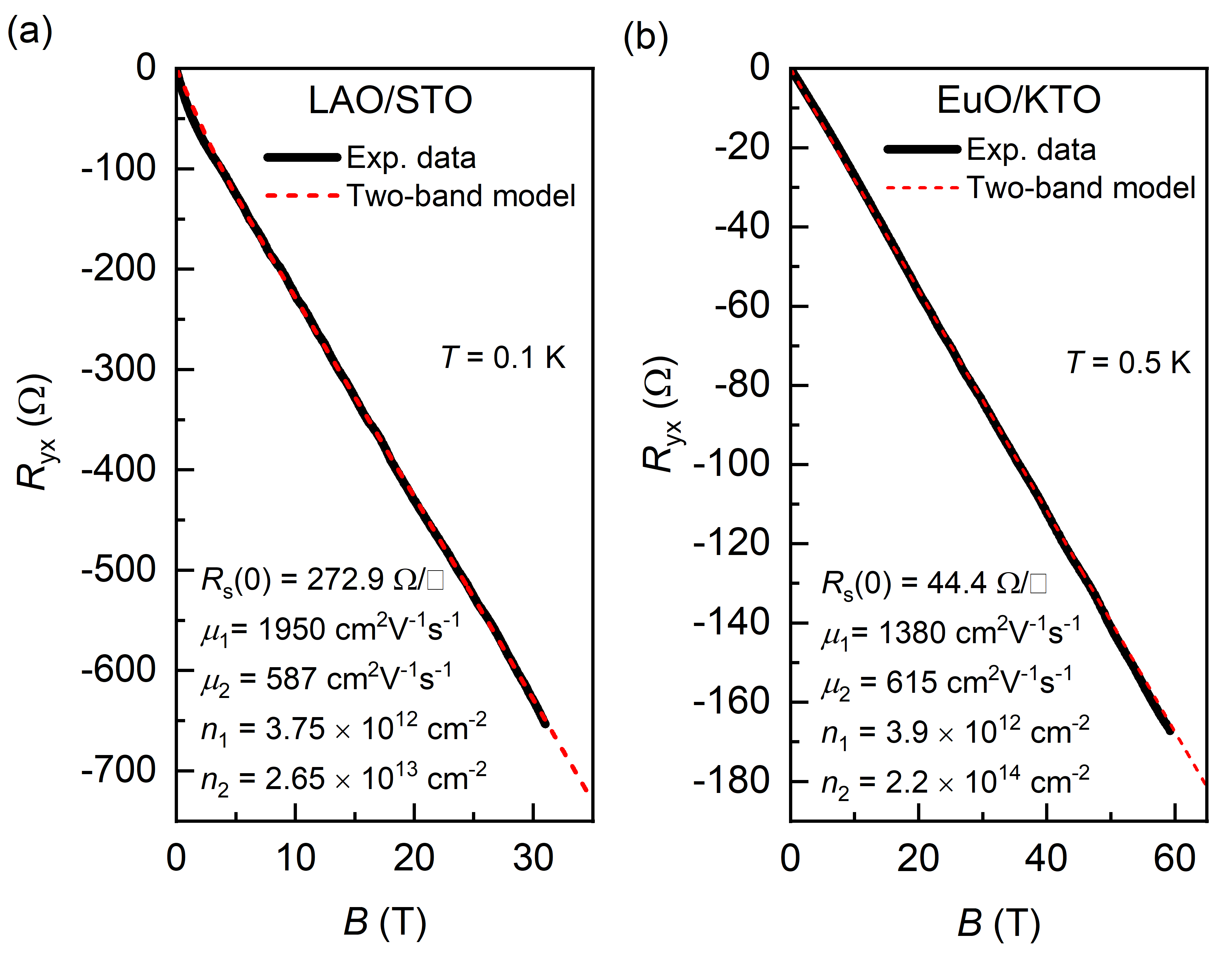}
\caption{\label{F2}{Hall resistance for (a) LAO/STO and (b) EuO/KTO fitted with the two-bands model.}}
\end{figure}

{From the two-band Drude model, the magnetic field dependence of $R_{yx}$ and the zero-field sheet resistance $R_s(0)$ are defined as}

\begin{equation}
{R_{yx}(B)=-\frac{B}{e}\frac{(n_1\mu_1^2+n_2\mu_2^2)+(\mu_1\mu_2B)^2(n_1+n_2)}{(n_1\mu_1+n_2\mu_2)^2+(\mu_1\mu_2B)^2(n_1+n_2)^2}, ~~~~\rm{and}~~~~}
\end{equation}

\begin{equation}
{R_s (0) = \frac{1}{e} \left(\frac{1}{n_1\mu_1+n_2\mu_2}\right),}
\end{equation}

{We fit the Hall resistance data for both LAO/STO and EuO/KTO interfaces to the Eq. (A6) in such a way that the fitting parameters $n_{1}$, $n_{2}$, $\mu_{1}$ and $\mu_{2}$ satisfy Eq. (A7) where $R_s(0)$ is experimental value of the sheet-resistance at zero-field. The fitting parameters along with  $R_s(0)$ are incorporated in Fig. A12. As one can see, the fitting parameters justify the simultaneous existence of low-mobility (high-density) and high-mobility (low-density) subbands. For both interfaces, the density of the high-mobility carriers is comparable to the SdH density.}


\begin{figure*}[!htp]
\includegraphics[width=7in]{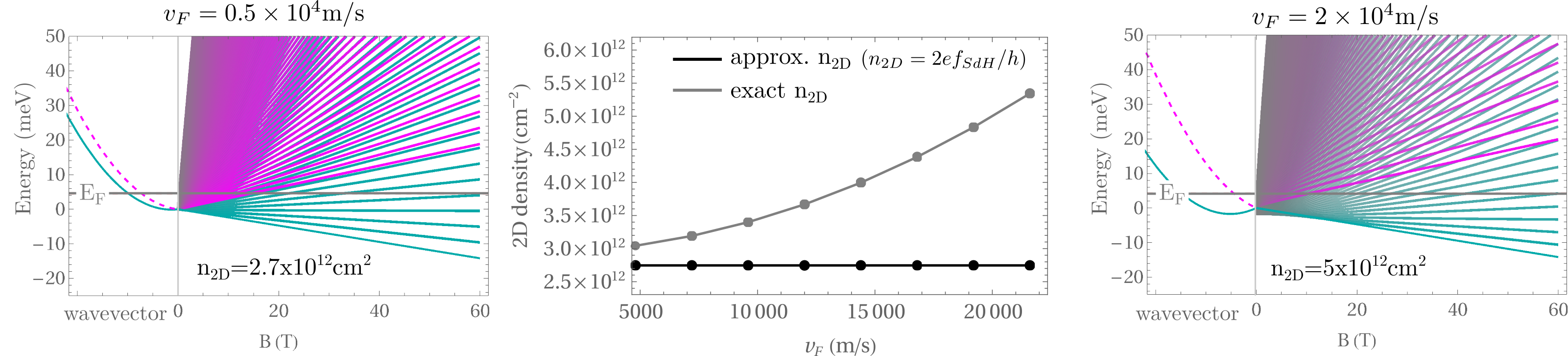}
\caption{\label{F2}{Theoretical calculation of SdH density for quadratic and quadratic+linear dispersion. Left and right panels are the electronic bands along with Landau levels for two different values of the Fermi velocity $v_f = 0.5 \times 10^4$ m/s and $2 \times 10^4$ m/s. The middle panel compares the density for models, quadratic (black symbols) and quadratic+linear (grey symbols), as a function of the Fermi velocity. The carrier density is estimated using the Eq. (A10)}}
\end{figure*}

\subsection{{Theoretical SdH oscillations}}
\label{app10}
{The numerical SdH oscillations presented in the main manuscript [Fig.~\ref{F7}] and in this Appendix (Fig.~\ref{F11}) is obtained by using that the differential magnetoresistivity $\delta R_{xx}\left(B\right)=\left[R_{xx}\left(B\right)-R_{xx}\left(B=0\right)\right])$, is proportional to the differential density of states (DOS) at the Fermi level ($\varepsilon_{F}$)
i.e., }
\begin{equation}
{\delta R_{xx}\left(B\right)\propto{g\left(\varepsilon,B\right)-g\left(\varepsilon_{F},B=0\right)}.}
\label{sdh-theory}
\end{equation}
{To simulate these oscillations, we will be following the framework developed in Refs.~\cite{candidoPRR2023,PhysRevB.109.115303,ferreira2022engineering}. Accordingly, we obtain the DOS assuming the LLs ($\varepsilon_{n}$  $n\in\mathbb{N}_{0}$) are broadened following a Lorenztian function with corresponding broadening $\Gamma$, i.e.,  }
\begin{equation}
{g\left(\varepsilon,B\right)=\frac{eB}{2\pi\hbar}\sum_{n}\delta\left(\varepsilon-\varepsilon_{n}\right)\rightarrow\frac{eB}{2\pi\hbar}\sum_{n}L_{\Gamma}\left(\varepsilon-\varepsilon_{n}\right),}
\end{equation}
{with $L_{\Gamma}\left(x\right)=\frac{1}{\pi}\frac{\Gamma/2}{x^{2}+\left(\Gamma/2\right)^{2}}$. The relationship between the $\varepsilon_{F}$ and 2D density of carrier ($n_{2D}$) is straightforward determined by solving}
\begin{align}
{n_{2D}} & {=\int_{-\infty}^{\infty}d\varepsilon g\left(\varepsilon,B\right)f_{FD}\left(\varepsilon-\varepsilon_{F}\right)},\nonumber \\
 & {=\frac{eB}{2\pi\hbar}\sum_{n}\int_{-\infty}^{\infty}d\varepsilon L_{\Gamma}\left(\varepsilon-\varepsilon_{n}\right)f_{FD}\left(\varepsilon-\varepsilon_{F}\right),}
\end{align}
{where $f_{FD}\left(\varepsilon-\varepsilon_{F}\right)$ is the Fermi-Dirac distribution. Once we determine $\varepsilon_{F}$ from $n_{2D}$, we are able to plot the SdH-oscillations using Eq.~(\ref{sdh-theory})}
\begin{equation}
{\delta R_{xx}\left(B\right)\propto\frac{eB}{2\pi\hbar}\sum_{n}L_{\Gamma}\left(\varepsilon_{F}-\varepsilon_{n}\right)}
\end{equation}
{where $\varepsilon_{n}$ are the LL energies originating from our Hamiltonian. In Fig.~\ref{F7} we plot the corresponding SdH-oscillations for three different models described by Hamiltonian Eq. (2): (a) pure quadratic Hamiltonian with Zeeman interaction ($m^*=1.5m_0$, $v_F=0$ and $g=9.5$); quadradic Hamiltonian with linear term and Zeeman interaction ($m^*=2m_0$, $v_F=2\times 10^4$~m/s and $g=15$; and (c) pure linear Hamiltonian with Zeeman interaction ($1/m^*\rightarrow 0$, $v_F=2\times 10^4$~m/s and $g=10$).}

\subsubsection{{Linear model}}

{In this subsection we show analytically that a linear model in the presence of Zeeman with corresponding Hamiltonian Eq. (2) host non-periodic SdH-oscillations with respect to $1/B$ and effective mass enhancement with $B$. Our starting point is a single basis Hamiltonian with spin up $\left(\uparrow\right)$
and down $\left(\downarrow\right)$, i.e., Hamiltonian Eq.(2) with $1/m^* \rightarrow 0$, }
\begin{equation}
{{\cal H}=\left(\begin{array}{cc}
\frac{1}{2}\mu_{B}gB & \hbar v_Fk_{+}\\
\hbar v_F k_{-} & -\frac{1}{2}\mu_{B}gB
\end{array}\right),}
\label{Hlinear}
\end{equation}
{To obtain the corresponding Landau levels, we use
the minimum coupling (Peierls substitution) $\hbar k_{i}\rightarrow\Pi_{i}=\hbar k_{i}-qA_{i}$,
with $\Pi_{x}=\frac{\hbar}{\sqrt{2}\ell_{B}}\left(a+a^{\dagger}\right)$,
$\Pi_{y}=\frac{\hbar}{i\sqrt{2}\ell_{B}}\left(a^{\dagger}-a\right)$,
and $\ell_{B}=\sqrt{\frac{\hbar}{eB}}$, yielding}
\begin{equation}
{{\cal H}=\left(\begin{array}{cc}
\frac{1}{2}\mu_{B}gB & \frac{v_F\sqrt{2}\hbar}{\ell_{B}}a^{\dagger}\\
\frac{v_F\sqrt{2}\hbar}{\ell_{B}}a & \frac{1}{2}\mu_{B}gB
\end{array}\right)=\left(\begin{array}{cc}
\Delta & v_{B}a^{\dagger}\\
v_{B}a & -\Delta
\end{array}\right),}
\end{equation}

{with $\Delta=\frac{1}{2}\mu_{B}gB$ and $v_{B}={\hbar v_F\sqrt{2}}/{\ell_{B}}$.
The corresponding LL energy levels are then straightforward calculated, and read}
\begin{equation}
{\varepsilon_{n}^{\pm}=\pm\sqrt{\Delta^{2}+v_{B}^{2}n}}
\end{equation}
{for $n\neq0$, and $\varepsilon_{n=0}^{\pm}=\pm\Delta$, with $+\left(-\right)$
representing $\uparrow\left(\downarrow\right)$. The analytical SdH-oscillations
in the resistivity are given by \cite{candidoPRR2023,PhysRevB.109.115303,ferreira2022engineering}}
\begin{equation}
{\delta R_{xx}\left(B\right)=\begin{subarray}{c}
\sum_{l=1}^{\infty}\end{subarray}R\left(T\right)e^{-l\pi\Gamma F'\left(\varepsilon_{F}\right)}\cos\left[2\pi lF\left(\varepsilon_{F}\right)\right]}
\end{equation}
{with F-function defined by}
\begin{equation}
{n \equiv F\left(\varepsilon\right) n =\frac{1}{v_{B}^{2}}\left(\varepsilon+\Delta\right)\left(\varepsilon-\Delta\right).}
\end{equation}
{Substituting $v_{B}={v_F\sqrt{2}\hbar}/{\ell_{B}}$ with $\ell_{B}=\sqrt{\hbar/eB}$
and $\Delta=\frac{1}{2}\mu_{B}gB$, we obtain }
\begin{equation}
{F\left(\varepsilon\right)=\frac{\varepsilon^{2}}{2ev^{2}\hbar}\frac{1}{B}-\frac{\left(\mu_{B}g/2\right)^{2}}{2ev^{2}\hbar}B.}
\end{equation}
{Therefore, it is evident that the linear model in the presence of Zeeman ($g\neq0$) will exhibit non-periodic oscillations with respect to $1/B$. The strength of the deviation from the $1/B$-periodic oscillations is dictated by the competition between Zeeman and the linear term via $\left(g/v\right)^{2}$. Furthermore, if we associate an effective mass to this model via $\varepsilon_{n+1}^{+}-\varepsilon_{n}^{+}=\hbar\omega_{c}^{*}$, we obtain for $\Delta\gg v_{B}\sqrt{n}$}
\begin{equation}
{\varepsilon_{n+1}^{+}-\varepsilon_{n}^{+}=\hbar\omega_{c}^{*}\approx\frac{1}{2}\frac{v_{B}^{2}}{\Delta},}
\end{equation}
{yielding the following effective mass enhancement with respect to the magnetic field}
\begin{equation}
{\frac{1}{m^{*}}=\frac{2v^{2}}{\mu g}\frac{1}{B}.}
\end{equation}
{Thus, we also show that an effective mass enhancement is also originated by the linear term in the presence of Zeeman interaction. }

\subsubsection{Higher $k$-terms}

\begin{figure}[!htp]
\includegraphics[width=2.5in]{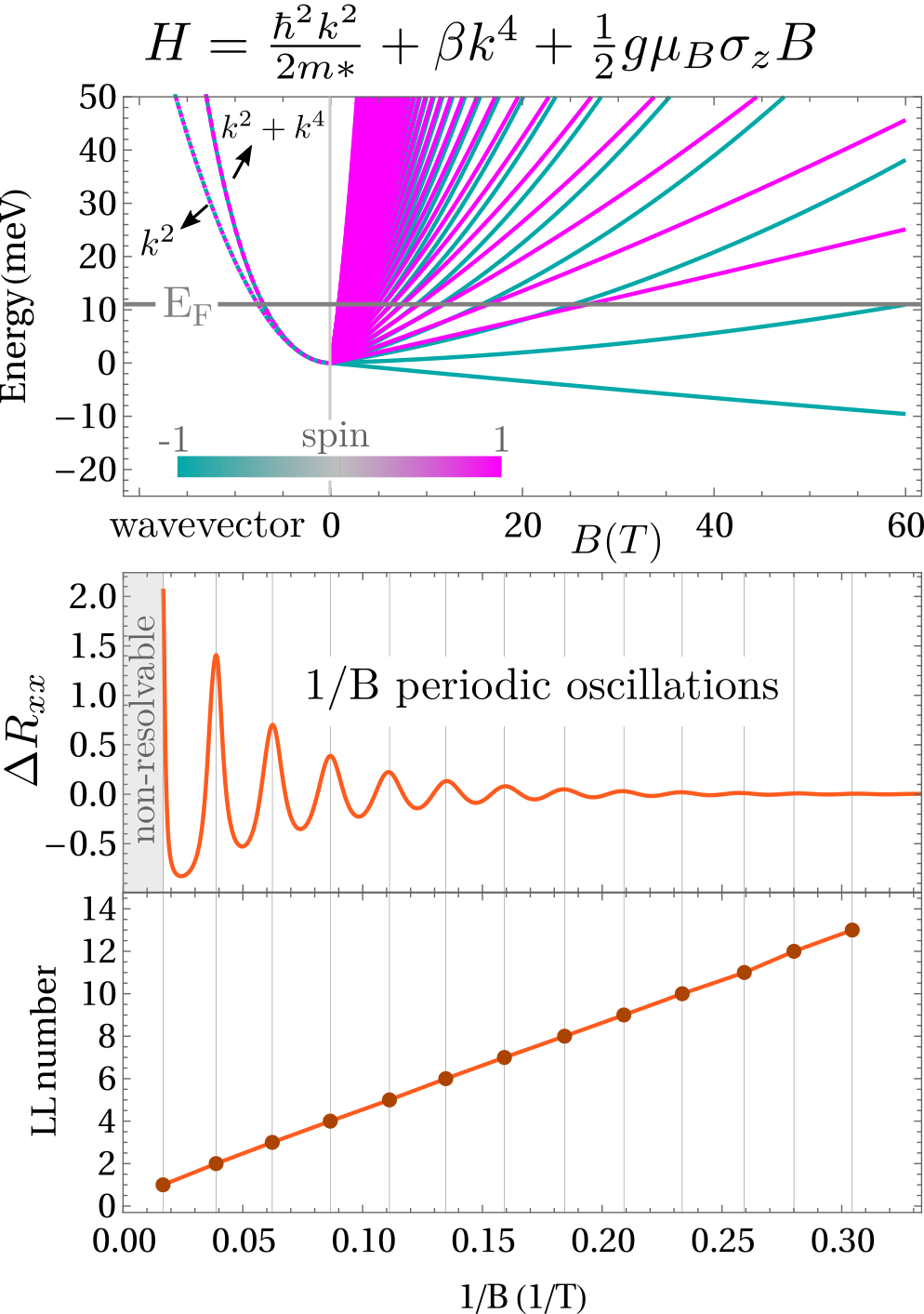}
\caption{\label{F11}{(a) Plot of the electronic band structure and corresponding Landau Levels for a 2DEG with quadratic and quartic terms in $k$, and with Zeeman interaction (g-factor), $m^*=0.5m_0$ , $g=10$ and $\beta=100$ meV nm$^4$. The lower panels shows the corresponding SdH-oscillations and LL index as a function of $1/B$.}}
\end{figure}

{Here we derive the SdH oscillations in the presence of higher order terms in k. To start with, we consider }
\begin{equation}
{{\cal H}=\frac{\hbar^{2}k^{2}}{2m^{*}}+\beta\left(k_{x}^{2}+k_{y}^{2}\right)^{2}+\frac{1}{2}g\mu_{B}B\sigma_z }
\end{equation}
{Using the minimal coupling, we obtain the Hamiltonian
\begin{align}
{\cal H} & =\hbar\omega_{c}\left(a^{\dagger}a+1/2\right)+\frac{1}{2}g\mu_{B}B\sigma_{z},\nonumber \\
&+\beta\left(\frac{2eB}{\hbar}\right)^{2}\left[\left(a^{\dagger}a\right)^{2}+a^{\dagger}a+1/4\right],
\end{align}}
{with corresponding energies given by 
\begin{align}
\varepsilon_{n}^{\pm} =\hbar\omega_{c}\left(n+1/2\right)+\hbar\omega_{Q}\left(n^{2}+n+1/4\right)\pm\frac{1}{2}g\mu_{B}B,\label{quarticLL}
\end{align}
with $\hbar\omega_{Q}=\beta\left(\frac{2eB}{\hbar}\right)^{2}$. Using $m^{*}/m_{0}=0.5$, $\beta=100\textrm{ meV nm}^{4}$ and $g=10$, in Fig.~\ref{F11} we plot the corresponding bulk bands vs wavevector, and LLs {[}Eq. (\ref{quarticLL}){]} as a function
of $B$. In the lower panels we plot the corresponding SdH-oscillations for $n_{2D}=2\times10^{12}\textrm{ cm}^{-2}$, where we observe periodic oscillations with respect to $1/B$. The LL index vs $1/B$ also confirm the periodicity.}

\bibliographystyle{apsrev4-2-Title}

%

\end{document}